\documentclass[twocolumn,reprint,superscriptaddress,nofootinbib]{revtex4-2}
\usepackage{graphicx}% Include figure files
\usepackage{dcolumn}% Align table columns on decimal point
\usepackage{bm}% bold math
\usepackage{epsfig}
\usepackage{epstopdf}
\usepackage{tabularx}
\usepackage{float}
\usepackage{booktabs}
\usepackage{hyperref}% add hypertext capabilities
\hypersetup{colorlinks=true, citecolor=blue, urlcolor=blue, linkcolor=blue}
\begin{document}
\title{
	Simultaneous calculation of elastic scattering, fusion, and direct cross sections for reactions of weakly-bound projectiles}
\author{H. M. Maridi}
\email[Corresponding author:
%Permanent address: Department of Physics, Faculty of Applied Science, Taiz University, Taiz, Yemen;
]{hasan.maridi@manchester.ac.uk}
\affiliation{Department of Physics and Astronomy, University of Manchester, Manchester M13 9PL, United Kingdom}
%\affiliation{Physics Department, Faculty of Applied Science, Taiz University, Taiz, Yemen}
\author{N. Keeley}
%\email{nicholas.keeley@ncbj.gov.pl}
\affiliation{National Centre for Nuclear Research, ul.\ Andrzeja So\l tana 7, 05-400 Otwock, Poland}
\author{K. Rusek}
%\email{rusek@slcj.uw.edu.pl}
\affiliation{Heavy Ion Laboratory, University of Warsaw, ul.\ Pasteura 5a, 02-093, Warsaw, Poland}
\begin{abstract}
Simultaneous analyses are performed of cross section data for elastic scattering, fusion, Coulomb breakup, and other direct yields for the $^{6}$He+$^{209}$Bi system at near Coulomb-barrier energies. The bare and dynamical polarization potentials are constructed microscopically from the structure of the colliding nuclei and they reproduce all the data well with only two adjustable parameters. This method of calculation can be successfully applied to the reactions of weakly-bound and exotic projectiles with heavy targets.
\end{abstract}
%\pacs{21.10.Gv,25.60.-t,25.60.Bx}% PACS, the Physics and Astronomy
                             % Classification Scheme.
%\keywords{one-neutron halo nuclei; $d\sigma (E1)/d\varepsilon$; Coulomb breakup; eikonal model; arXiv:}

\date{\today}%
\maketitle
%=========================================================

\section{\label{sec:int} Introduction}
%==========================================================
In recent decades there has been significant progress in the exploration of the mechanisms involved in heavy ion collisions at energies close to the Coulomb barrier. One particular focus has been on reactions caused by radioactive halo nuclei, which may be visualized as consisting of one or two loosely-bound valence nucleons orbiting at a large distance from a tightly-bound core. When these reactions occur at low energies, close to the Coulomb barrier, they are mainly dominated by fusion and direct reactions like transfer and breakup. Reviews of the reactions induced by these exotic nuclei interacting with heavy targets may be found in Refs.~\cite{Kee09,Can15,Kol16}.

The optical model potential is often used to describe elastic scattering data within a single channel approach, and for heavy ion projectiles is most commonly of Woods-Saxon volume form for both real and imaginary parts. However, the imaginary potential is occasionally split into volume and surface components, with a short range volume term arranged to simulate the ingoing-wave boundary condition to model loss of flux due to fusion and a surface term with a longer range to account for loss of flux due to non-elastic direct reaction channels.
This is the so-called extended optical model introduced by Udagawa and collaborators \cite{Uda84,Uda85,Uda89,Hon89,Kim90} which gives good simultaneous fits to the fusion and elastic scattering data for a large variety of systems. The direct component can also be taken as a complex potential, i.e.\ including a real part, within this model, as in Refs.~\cite{Kim02a,Kim02b,So07,So10}.

In the reactions of weakly-bound projectiles with heavy targets the projectile can become polarized by and/or break up in the strong electric field of the target. The resulting strong Coulomb dipole excitation and breakup can be treated by introducing an additional interaction which influences the elastic scattering. This additional interaction is often referred to as the Coulomb dynamical polarization potential (CDPP).
Recently \cite{Mar21,Mar22}, a new expression for the CDPP was obtained by solving the Schr\"odinger equation for the internal motion of an exotic neutron-rich projectile (considered as a two-body deuteronlike cluster structure) incident on a heavy target nucleus using the adiabatic approximation.
However, in the optical model (OM), the CDPP potential alone cannot entirely account for the long-range interactions in these exotic systems. To address this issue, a long-range nuclear dynamical polarization potential (NDPP) is introduced to account for the nuclear breakup and transfer reactions, so that the direct surface potential now consists of the CDPP plus the NDPP. This NDPP utilizes either a volume or surface Woods-Saxon type potential, usually with large radius and/or diffuseness parameters, see for example Refs.~\cite{So14,So15a,San08}.

In this work we present a form of the extended optical-model potential which is able simultaneously to reproduce elastic scattering, fusion, Coulomb breakup, other direct yields, and total reaction cross section data with two adjustable parameters. We apply this potential to calculations for the $^{6}$He+$^{209}$Bi system.
%=========================================================

\section{\label{sec:theory} Theory}
%==========================================================
%=========================================================

\subsection{\label{sec:OP} The optical potentials}
%==========================================================
%=========================================================

\subsubsection{\label{sec:bare} Bare nuclear potential}
%==========================================================
The nuclear interaction in the absence of coupling effects is represented by a short-range complex nuclear potential---the ``bare potential''---which is often of volume Woods-Saxon form.
However, in this work the real part of the bare nuclear potential is taken from the S\~{a}o Paulo potential (SPP) \cite{Cha02}, which reproduces with reasonable accuracy the experimental angular distributions of a large number of stable systems over a wide energy range with no adjustable parameters \cite{Alv21,Alv03}. It is obtained by multiplying the double folding potential
($V_F$) by an energy dependent factor:
\begin{equation}
	\label{eq:spp1}
	V_{\mathrm{SPP}} ({\bf R})=V_F ({\bf R})\mathrm{e}^{-4 \upsilon^2 /c^2},
\end{equation}
where $\upsilon$ is the relative velocity between the projectile and target, $c$ is the speed of light and $V_F$ is given by \cite{Sat79}
\begin{equation}
	\label{eq:vfd1}
	V_F ({\bf R})= \int{\rho_p(\textbf{r}_p) \rho_t(\textbf{r}_t) \upsilon_{nn}(s)d\textbf{r}_p d\textbf{r}_t},
\end{equation}
where $\rho_p(\textbf{r}_p), \rho_t(\textbf{r}_t)$ are the nuclear matter density distributions for projectile and target nuclei, respectively, $s=|\textbf{R}-\textbf{r}_p+\textbf{r}_t|$ is the distance between the two nucleons, and $\upsilon_{nn}(s)$ is the effective $NN$ interaction, which in this case is the zero-range effective \emph{NN} interaction $\upsilon_{nn}(s)=V_0 \delta(s)$, with $V_0=-456$ MeV \cite{Cha02}.

The imaginary potential may be of Woods-Saxon form or the SPP potential multiplied by a normalization factor \cite{Alv03}, in which case the bare optical potential is given by
%$U_0 (R)= (1 + 0.78 i)V_F(R)$
\begin{eqnarray}
	\label{eq:OP3}
	U_{N} (R)=N_{R} V_{\mathrm{SPP}} (R)+iN_{I} V_{\mathrm{SPP}} (R)
\end{eqnarray}
where $N_R$ and $N_I$ are the normalization factors that fit the data and simulate the polarization effects. 
A systematic analysis of many stable tightly bound nuclei in Ref.~\cite{Alv03} arrived at $N_R = 1.00$ and $N_I = 0.78$ as reference values of the normalization factors, and this bare potential was used recently to analyze reactions with exotic and stable nuclei \cite{Alv21}.
The SPP has also been shown to be a reasonable basis for the analysis of fusion reactions induced by stable weakly-bound nuclei \cite{Cre07}.

Note that since the weakly-bound nuclei considered here are composed of a core nucleus and one or two valence nucleons, the short-range bare potential may also be obtained by fitting suitable projectile core-target elastic scattering data.
%=========================================================

\subsubsection{\label{sec:CDPP} Coulomb dynamical polarization potential (CDPP)}
%=========================================================
Recently \cite{Mar21,Mar22}, the CDPP was obtained by solving the formalism for the scattering of a weakly-bound two-body projectile consisting of a core plus a cluster of $n$ valence neutrons from a heavy target.
To solve the Schr\"{o}dinger equation of the system and obtain the CDPP one may use the adiabatic approximation $\Psi ({\bf r,R})\approx \psi ({\bf R}) \phi ({\bf r,R})$, where $\psi ({\bf R})$ refers to the wave function of the center of mass and $\phi ({\bf r,R})$ to that of the internal motion of the projectile; ${\bf R}$ and ${\bf r}$ are the coordinates of the projectile-target and the projectile valence-core systems, respectively.
The resultant CDPP $\delta U_{C}$ must obey
\begin{equation}
\label{eq:dvR4}
\left(\frac{\varepsilon_0+\delta U_{C} (R)}{\varepsilon_0}\right)  H_0^{+}(\rho)F_{0}(\rho) -Q^2(R) H_0^{+'}(\rho)F_0^{'}(\rho)= Q(R),
\end{equation}
where $H_0^{+}=G_0+iF_0$, with $F_0$ and $G_0$ the regular and irregular Coulomb functions in $\rho= k(R)R$ and the Sommerfeld parameter $\eta=(m_{c}^2/\mu_p)Z_P Z_T e^2 /{\hbar}^2 k(R)$, where $\mu_p$ is the core-valence reduced mass, $m_{c}$ the mass of the charged core, and $Z_P$ and $Z_T$ the charges of the projectile and target, respectively. $Q(R) =\frac{\mu_p}{m_{c}} \frac{k(R)} {\kappa_0}$ with $\kappa_0 = \sqrt{-2\mu_p\varepsilon_0/\hbar^2}$, where
$\varepsilon_0$ is the binding energy of the valence neutron or neutron cluster with respect to the charged core of the projectile. If the core is in an excited state $\varepsilon_0$ is replaced by $\varepsilon^*_0=\varepsilon_0+\varepsilon_{I^{\pi}_c}$ where $\varepsilon_{I^{\pi}_c}$ is the excitation energy of the core state of spin-parity $I^{\pi}_c$.
The wave number of the charged core in the field of the target that is associated with the wave function of the internal motion of the projectile,
\begin{equation}
	\label{eq:k}
k(R) = {\sqrt{\frac{2 m_{c}^2}{\mu_p {\hbar}^2}(V_C ({ R})+\varepsilon_0 +\delta U_{C} (R))}},
\end{equation}
may be approximated as in Ref.\ \cite{Mar21} by assuming $\delta U_{C} (R) << V_C ({ R})$ to obtain analytical formulas for
the real and imaginary parts of the CDPP:
\begin{eqnarray}
\label{eq:CDPP1}
\delta V_{C} (R)&=& \varepsilon_0 \left[ \frac{QG_{0}F_{0}+Q^2 G_{0}F_{0}G_{0}'F_{0}' +Q^2 F_{0}^2 F_{0}'^2}{F_{0}^4 + G_{0}^2 F_{0}^2}-1 \right] \nonumber \\
\delta W_{C} (R)&=& \varepsilon_0 \left[ \frac{Q^2 F_{0}F_{0}' -Q F_{0}^2}{F_{0}^4 + G_{0}^2 F_{0}^2} \right].
\end{eqnarray}
Note that $k(R)$ depends parametrically on the Coulomb potential between the projectile and target, $V_C ({ R})$ and is different from the wave number of the center-of-mass motion of the system which describes the motion of the projectile along the Rutherford trajectory, $K=\sqrt{2 \mu (E-\varepsilon_0) / \hbar^2 }$ where $E$ is the incident energy of the projectile and $\mu$ is the reduced mass of the projectile-target system.
The CDPP of Eq.~(\ref{eq:CDPP1}) thus depends on the structure of the system but not on the incident energy of the projectile.
At this point we should add that the local momentum $k(R)$ becomes complex at large distances where $R > Z_P Z_T e^2/\varepsilon_0$, and in this case we would need to use the Coulomb function for the complex arguments $\rho$ and $\eta$. However, in this work, for the $^{6}$He+$^{209}$Bi reaction using $\varepsilon_0=1.6$ MeV for $^{6}$He, $k(R)$ becomes complex when $R>149.4$ fm and this is larger than the effective distances needed to analyze this reaction.   

This CDPP has been successfully applied to the calculation of Coulomb dissociation at high energies by including excitation to the continuum \cite{Mar22}. In addition, the dipole polarizability, $\alpha_{0}$, may be obtained by fitting the long range real part of Eq.~(\ref{eq:CDPP1}) by the classical expression $\delta V_C = -\frac{1}{2} \alpha_0 \frac{Z_T^2 e^2}{R^4}$ at large distances \cite{Mar21}:
	\begin{equation}
	\label{eq:CDPPR}
	%\alpha_{0}  & = &\frac{\sqrt{2} \pi^{2}}{16}  max
	\delta V_C (R)  \longrightarrow  -\frac{1}{2}\left(  {\hbar^{2} \mu_p Z_P^2 e^2 \over 16 m_c^2 \varepsilon_0^{2} } \right)  \frac{ Z_T^2 e^2}{R^4} 
\end{equation}
so that
\begin{equation}
	\label{eq:alp0}
	\alpha_{0} =\frac{1}{16} {\hbar^{2} \mu_p Z_P^2 e^2 \over m_c^2 \varepsilon_0^{2} }.
\end{equation}
The resulting values are in good agreement with those obtained using other methods \cite{Mar21}.  
The same expression for the polarizability can be derived from the response function $dB(E1,\varepsilon)/ d\varepsilon$ obtained in the simple cluster model \cite{Ber92}
\begin{eqnarray}
	{dB(E1,\varepsilon) \over d\varepsilon}
	= {3 \hbar^{2} \mu \over \pi^{2} m_c^2 } {\sqrt{\varepsilon_0} (\varepsilon-\varepsilon_0)^{3/2}\over \varepsilon^4}
\end{eqnarray}
from which we get
\begin{equation}
	\alpha_{0} =\frac{8 \pi}{9} \int_{\varepsilon_0}^\infty \frac{1}{\varepsilon}\frac{dB(E1)}{d\varepsilon}d\varepsilon=\frac{1}{16} {\hbar^{2} \mu_p Z_P^2 e^2 \over m_c^2 \varepsilon_0^{2} },
\end{equation}
where the integrated $B(E1)$ is given by
\begin{equation}
	B(E1) =\int_{\varepsilon_0}^\infty \frac{dB(E1)}{d\varepsilon}d\varepsilon=\frac{3}{16 \pi} {\hbar^{2} \mu_p Z_P^2 e^2 \over m_c^2 \varepsilon_0}=\frac{3}{\pi} \alpha_{0} \varepsilon_0.
\end{equation}
The CDPP is thus seen to be indirectly dependent on $dB(E1,\varepsilon)/ d\varepsilon$.

To examine the behavior of the CDPP at large separation energies we use the second order Wentzel-Kramers-Brillouin (WKB) approximation for the Coulomb functions in Eq.~(\ref{eq:dvR4}). As $\varepsilon_0$ is increased $k(R)$, $Q(R)$, and $\rho=k(R)R$ decrease whereas $\eta$ increases. Thus, for large values of $\varepsilon_0$ we may use the approximations of the Coulomb functions for $2\eta >> \rho$ \cite{Abr72},
\begin{eqnarray}
	F_{0}&\simeq& \frac{\beta}{2} e^{\gamma}, \ \ \ F_{0}'\simeq \left( \beta^{-2}+\frac{1}{8\eta}\frac{\beta^{4}}{t^2}\right) F_{0}, \nonumber \\
	G_{0}&\simeq&\beta e^{-\gamma}, \ \ \  G_{0}'\simeq-\left( -\beta^{-2}+\frac{1}{8\eta}\frac{\beta^{4}}{t^2}\right) G_{0}
\end{eqnarray}
where
\begin{eqnarray}
	\label{eq:CF1}
	t&=& \frac{\rho}{2\eta}, \ \  \beta= \left( \frac{t}{t-1}\right)^{1/4},\nonumber \\
	\gamma&=&2\eta \left( \sqrt{t(1-t)}+\textrm{arcsin} \sqrt{t}-\frac{\pi}{2}\right). 
\end{eqnarray}
Similarly to Ref.~\cite{Bor07}, by expressing $k(R)$, $\rho$, $\eta$, and $Q(R)$ as functions of $\delta U_{C} (R)$, $\varepsilon_0$, and $R$ and expanding in the small parameter $\delta U_{C} (R)/\varepsilon_0$, one finds \cite{Bor07}
\begin{eqnarray}
	\label{eq:CDPP2}
	\delta V_{C} (R)&=& -\frac{1}{2} \alpha_0  \frac{ Z_T^2 e^2}{R^4} \nonumber \\
	\delta W_{C} (R)&=& -\left[ \sqrt{\frac{1}{2} \alpha_0  \frac{ Z_T^2 e^2}{R^4} \varepsilon_0} +\frac{1}{4}\alpha_0  \frac{ Z_T^2 e^2}{R^4} \right] e^{2\gamma}
\end{eqnarray}
where $\alpha_0$ is given by Eq.~(\ref{eq:alp0}). Since $e^{2\gamma}$ is small, the imaginary CDPP tends to zero for large $\varepsilon_0$. In addition, this result is expected from Eq.~(\ref{eq:CDPP1}) where for small $\rho$, $F(\rho)$ and $F'(\rho)$ are also small but $G(\rho)$ and $G'(\rho)$ have very large values and $\delta W_{C} (R) \longrightarrow 0$.

\begin{figure}
		\centering
		\includegraphics[width=0.45\textwidth,clip=]{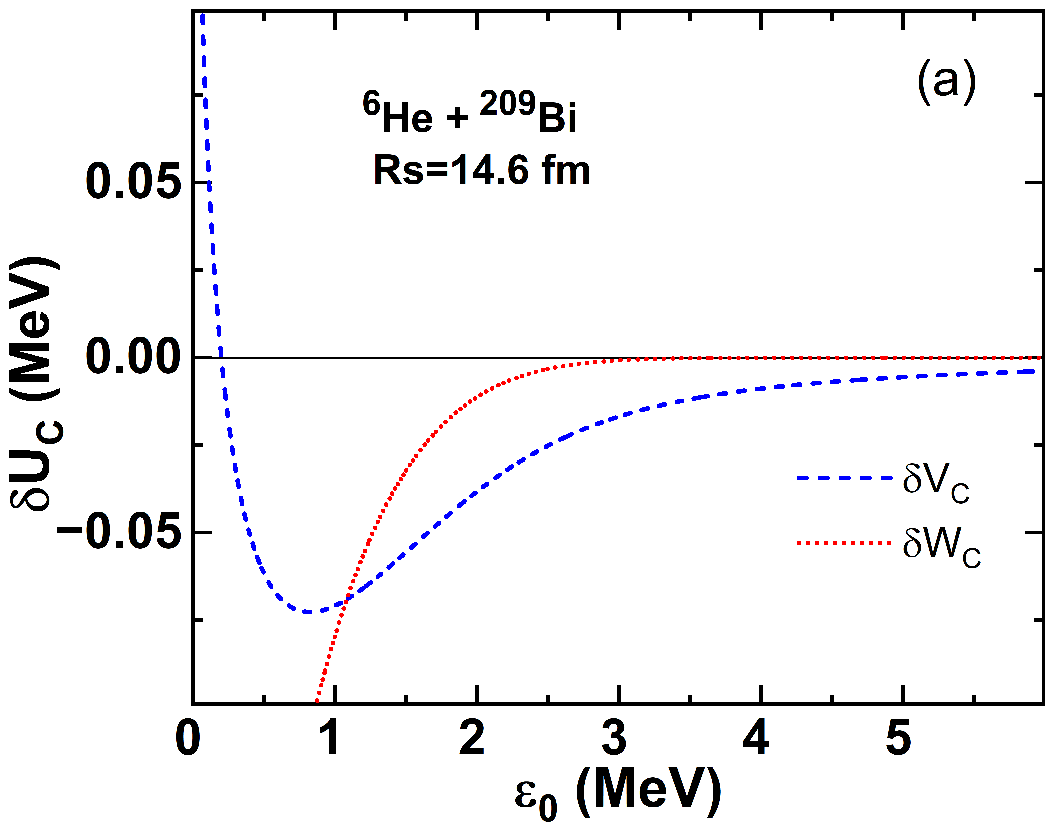}
		\includegraphics[width=0.45\textwidth,clip=]{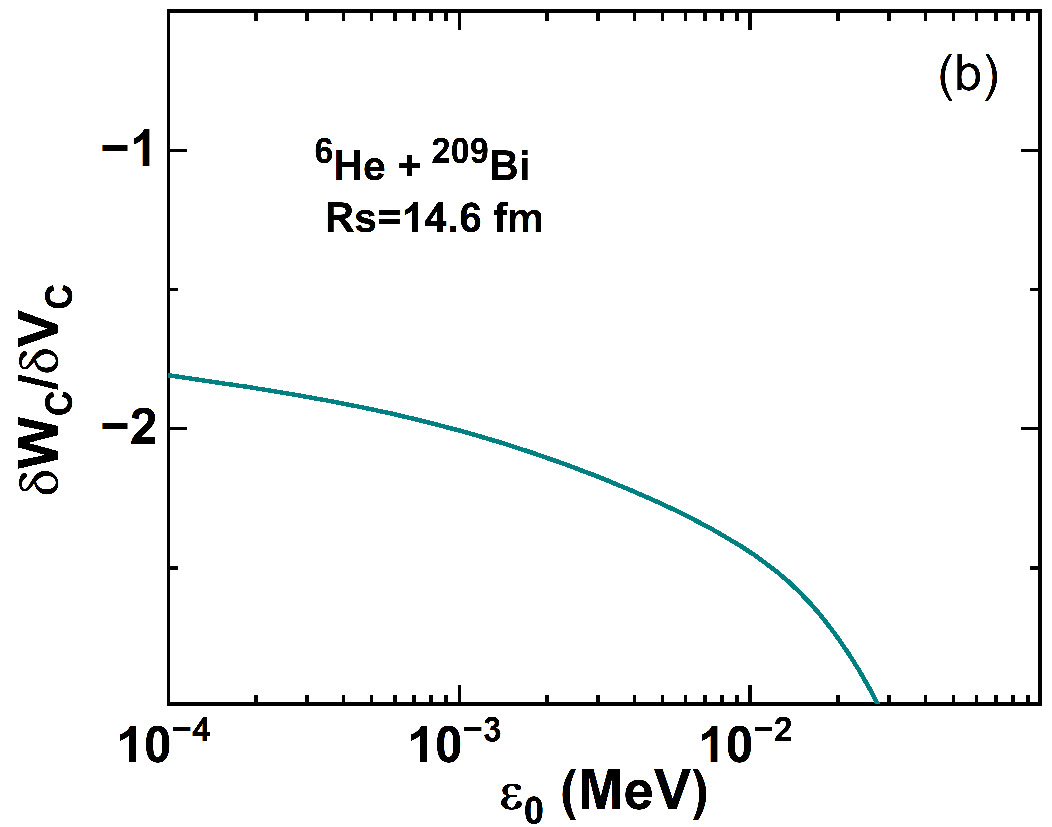}
		\caption{\label{fig:WV} The behavior of the CDPP (\ref{eq:CDPP1}) for $^{6}$He+$^{209}$Bi at the sensitivity radius $R_s=14.6$ fm  
                 as a function of the separation energy. The dotted and dashed curves in (a) represent the real and imaginary parts of the CDPP,
                 respectively and the solid line in (b) denotes the imaginary-to-real ratio. Note the logarithmic separation energy scale in (b).}
	\end{figure}

For small values of $\varepsilon_0$, $2\eta\sim \rho$ and $Q(R)>>0$. For large enough values of $\eta$ we may use the approximation where the Coulomb functions can be represented by the Airy functions \cite{Abr72}
% $F(\rho)$ and $F'(\rho)$
\begin{eqnarray}
	G_{0}+F_{0}&\sim& \pi^{1/2} (2\eta)^{1/6}\left[ Bi(x)+iAi(x)\right], \nonumber \\
	G_{0}'+iF_{0}'&\sim&-\pi^{1/2} (2\eta)^{-1/6}\left[ Bi'(x)+iAi'(x)\right]
\end{eqnarray}
where $x=\left(2\eta-\rho \right) \left(2\eta \right)^{-1/3}$. 
As $\varepsilon_0\longrightarrow 0$, $\rho=2\eta>>0$ and these expressions become
\begin{eqnarray}
	\label{eq:CF12}
	G_{0}+iF_{0}&\sim& \frac{\Gamma(1/3)\sqrt{w}}{2\sqrt{\pi}}\left(\sqrt{3}+i \right), \nonumber \\
	G_{0}'+iF_{0}'&\sim& \frac{\Gamma(2/3)}{2\sqrt{\pi w}} \left(-\sqrt{3}+i \right), 
\end{eqnarray}
where $w=\left(\frac{2\eta}{3}\right)^{1/3}$. From Eq. (\ref{eq:dvR4}) one then obtains
\begin{eqnarray}
	\label{eq:CDPP3}
		\delta V_{C} (R)+i\delta W_{C} (R)\sim \frac{\varepsilon_0 Q(R)^2}{2w^2} \frac{\Gamma(2/3)^2}{\Gamma(1/3)^2} \left( -1+i\sqrt{3}\right) \nonumber \\
			\sim 0.1548\left[ \frac{1}{2}\frac{\mu_p \hbar^2 (Z_P Z_T e^2)^2}{16m_c^2 R^4}\right] ^{1/3} \left( 1-i\sqrt{3}\right). \nonumber \\
\end{eqnarray}
Thus, while for large breakup energies the CDPP becomes purely real as in the adiabatic expression, conversely, when the breakup energy tends to zero we get a small repulsive real part and an attractive imaginary part with $\left| \delta W_{C} (R) / \delta V_{C} (R) \right| \longrightarrow \sqrt{3}$, independent of $\varepsilon_0$.   
%===============
To check these limits, Fig. \ref{fig:WV} presents the real and imaginary CDPPs from Eq.~(\ref{eq:CDPP1}) and their ratio $\delta W_C/\delta V_C$ for $^{6}$He+$^{209}$Bi at the sensitivity radius $R_s=14.6$ fm \cite{Gar07} as functions of the separation energy ($\varepsilon_0$). It is clear that the calculated real and imaginary CDPPs and their ratio behave as expected from the above approximations at small and large separation energies. 	

%===============
This behavior is similar to that of the dynamic polarization potential induced by dipole Coulomb excitation to a distribution of dipole states introduced in Refs.~\cite{And95,And94}:
\begin{eqnarray} \label{potential_coulomb-inela}
	\delta U_{C} (R) &=& \frac{4 \pi}{9} \frac{Z^{2}_{t} e^{2}}{\hbar v} \frac{B(E1)}{(R-a_{0})^{2} R}
	\nonumber \\
	&\times& ~\left[ g\left( \frac{R}{a_{0}}-1,\xi\right) + i f\left( \frac{R}{a_{0}}-1,\xi\right) \right] ,
\end{eqnarray}
with
\begin{eqnarray}
	f\left( \frac{R}{a_{0}}-1,\xi\right) &=&4\xi^{2}\left( \frac{R}{a_{0}}-1 \right)^{2} e^{-\pi \xi} K''_{2i\xi}\left[ 2\xi\left( \frac{R}{a_{0}}-1\right) \right] , \nonumber\\
	g\left( \frac{R}{a_{0}}-1,\xi\right)  &=& \frac{P}{\pi} \int^{\infty}_{-\infty} \frac{f\left( \frac{R}{a_{0}}-1,\xi\right) }{\xi-\xi'}  d\xi' 
\end{eqnarray}
where $a_{0}$ is the distance of closest approach in a head-on collision, $K''$ is the second derivative of the modified Bessel function and $\xi$ = $a_{0} \varepsilon_0$/$\hbar v$ is the Coulomb adiabaticity parameter~\cite{And94}.
If $\xi$ is large, $f(z,\xi)\longrightarrow 0$ and $g(z,\xi) \longrightarrow z^2 / (z+1)^3 \xi$, and one thus obtains the adiabatic expression for the polarization potential which is purely real, as in Eq.~(\ref{eq:CDPPR}),
If $\xi$ is very small, then $f(z,\xi)\longrightarrow 1$ and $g(z,\xi) \longrightarrow 0$. The polarization potential then becomes purely imaginary.
%=========================================================

\subsubsection{\label{sec:NDPP} Nuclear dynamical polarization potential (NDPP)}
%==========================================================
However, the CDPP is usually insufficient completely to explain the long-range interactions in exotic systems. To tackle this problem, the direct polarization potential should include both the CDPP and a long-range nuclear dynamical polarization potential (NDPP) to factor in nuclear breakup and transfer. This NDPP typically employs a volume, for example Ref.~\cite{San08},  or surface, for example Refs.~\cite{So14,So15a,Hus94}, Woods-Saxon type imaginary potential, characterized by large radius and/or diffuseness parameters.
It is sometimes referred to as the direct potential and may also include a real part, see Refs.~\cite{Hus94,Kim02a,Kim02b,So07,So10}.

The dispersion relation, a consequence of causality, may be used to connect the real and imaginary parts of the NDPP according to
the following expression \cite{Nag85,Mah86} 
	\begin{equation}
		\bigtriangleup V(r;E)=\frac{P}{\pi} \int \frac{W(r;E')}{E'-E}dE'
	\end{equation}
where $P$ denotes the principal value of the integral.
However, the empirical optical potential need not necessarily satisfy the dispersion relation \cite{Nag85} and for weakly bound 
nuclei the coupling to breakup produces a large repulsive contribution to the real part of the optical potential, leading to the 
suggestion that the dispersion relation may not apply in these circumstances \cite{Mah86}. Indeed, the empirical optical potentials 
extracted from fits to near-barrier
$^6$Li + $^{208}$Pb elastic scattering data clearly do not satisfy such a dispersion relation, the surface strength of the real potential  
being essentially energy independent \cite{Kee94} while that of the imaginary potential {\it increases} as the incident energy is reduced
towards the barrier. Conversely, the $^7$Li + $^{208}$Pb potentials are broadly consistent with a dispersion relation \cite{Kee94}.
	
For $^{6}$He+$^{209}$Bi	the situation is less clear; while it is possible to fit the data with optical potentials which satisfy 
the dispersion relation within the uncertainties, see Refs.~\cite{Gar07,Gar08}, the relatively sparse angular distributions and 
the associated uncertainties make it hard to draw a definitive conclusion.
While more complete angular distributions are available for the near-barrier elastic scattering of the $^{6}$He+$^{208}$Pb system a similar
result was obtained; the best-fit optical potentials are consistent with the dispersion relation but the uncertainties
remain such that they are also broadly consistent with little or no energy dependence of the NDPP \cite{San08,Fer10}. In fact, good fits to 
the $^6$He + $^{208}$Pb elastic scattering data for incident energies of 16, 18 and 22 MeV are possible with a completely energy-independent
real potential and a long-range imaginary potential of fixed geometry and energy-dependent strength \cite{Kee19}. Effective potentials obtained
from continuum discretized coupled channel (CDCC) calculations, either four-body \cite{Fer10} or three-body \cite{Mac09}, do show some
energy dependence but its relative importance varies significantly with the radial region examined. Since the radial regions of sensitivity
to the potential are different for the real and imaginary parts \cite{Fer10} it is not clear whether these effective potentials will
satisfy a dispersion relation. 
%Given these uncertainties we have chosen to include only an imaginary NDPP since it is at least clear that this is the most important term, the main contribution to the real part of the total dynamical polarization potential coming from the CDPP. 

To fix the form of the NDPP we turn to semiclassical theory. Within this framework
the breakup probability is found to depend on the exponential $\exp(-b/a)$ \cite{Bon02}, where $b$ is the impact parameter and 
$a=1/(2\kappa_0)$ with $\kappa_0=\sqrt{2\mu \varepsilon_0 / \hbar^2 }$ the decay length of the initial wave and $\varepsilon_0$ the separation energy. A similar exponential form was assumed at large distances for the imaginary surface potential,
$W(R)\approx \mathrm{e}^{-(R-R_s)/a}$,  
which accounts for peripheral reactions like transfer and nuclear breakup. The strong absorption radius is taken as $R_s=1.4 (A_p^{1/3}+A_t^{1/3})$.

Note that a long-range surface Woods-Saxon potential with radius $R_L$ and diffuseness $a_L$ can be approximated by the exponential form at large distances \cite{Hus94,Has09}:
\begin{eqnarray}
\label{eq:DPP0}
\frac{\exp({R-R_L \over a_L})}{[1+\exp( {R-R_L \over a_L})]^2} \rightarrow \exp\left(-{R-R_L \over a_L}\right)
\end{eqnarray}
which is similar to the semiclassical formula with the same radius and diffuseness. 
The same applies to the volume Woods-Saxon shape:
\begin{equation}
\frac{1}{1+\exp( {R-R_L \over a_L})} \rightarrow \exp\left(-{R-R_L \over a_L}\right)
\end{equation}
so that using either form we can fix the radius and diffuseness from the semiclassical theory and just vary the strength.
In this work the long-range nuclear dynamical polarization potential is thus taken to be of derivative Woods-Saxon shape:
\begin{eqnarray}
\label{eq:DPP1}
\delta U_{N}&=&\delta V_{N}+i\delta W_{N} %&\equiv& V_{L}(R)+iW_{L}(R)
\nonumber\\&=&{-4 (V_L + iW_L)}\frac{\exp({R-R_L \over a_L})}{[1+\exp( {R-R_L \over a_L})]^2},
\end{eqnarray}
where $R_L=1.4 (A_p^{1/3}+A_t^{1/3})$ is the strong absorption radius and  $a_L=1/(2\kappa_0)$ the diffuseness, where $\kappa_0=\sqrt{2\mu \varepsilon_0 / \hbar^2 }$ and $\varepsilon_0$ is the separation energy. $V_L$ and $W_L$ are varied to fit the data.
%$r_{0L}=1.43-1.44$ fm \cite{So14}, 1.5 fm \cite{So15a},
For $^{6}$He: $a_L=2.0$ fm using $\varepsilon=0.975$ MeV (the actual $2n$ separation energy) and $a_L=1.565$ fm using $\varepsilon=1.6$ MeV (the ``effective'' separation energy used in the improved two-body cluster model of Moro {\it et al.\/} \cite{Mor07}), which may be compared with 1.25 fm ($^{6}$He+$^{209}$Bi \cite{Kim02a}), 2.29 fm ($^{6}$He+$^{208}$Pb) \cite{So14}, and 1.45 fm ($^{6}$He+$^{208}$Pb \cite{San08}) obtained empirically from fitting data.
For $^{11}$Li, $a_L=2.94$ fm which may be compared with 3.42 and 4.00 fm from $^{11}$Li+$^{208}$Pb \cite{So14}.
For $^{11}$Be, $a_L=3.38$ fm which may be compared with 3.50 fm from $^{11}$Be+$^{64}$Zn \cite{Pie10,Pie12} and 3.2 fm from $^{11}$Be+$^{64}$Zn \cite{So15a}. This therefore seems a reasonable basis for fixing the geometry of the NDPP. 
%=========================================================

\subsubsection{\label{sec:total} Total optical potential}
%==========================================================
The polarization potentials are added to the ``bare'' optical potential to give the generalized optical potential.
According to the Feshbach theory \cite{Fes58}, the effective optical potential can be written as $U_{N}+\delta U$ where $\delta U \equiv U_{\mathrm{pol}}(R)$ is the dynamical polarization potential. Here we have Coulomb and nuclear contributions.
The total projectile-target optical potential is given as:
\begin{eqnarray}
\label{eq:OP1}
U_{\mathrm{OP}}(R)=U_C(R)+U_{N} (R) + \delta U_{C} (R) +\delta U_{N} (R)
\end{eqnarray}
where $U_C(R)=V_C(R)$ is the usual real Coulomb potential with a radius of $R_C = 1.25 ({A_p}^{1/3} + {A_t}^{1/3})$, $U_{N} (R)$ the bare nuclear potential that accounts for the fusion, and $\delta U_{C} (R)=\delta V_{C} (R) +i\delta W_{C} (R)$ is the CDPP (\ref{eq:CDPP1}) that represents the dipole polarization and Coulomb breakup. $\delta U_{N} (R)=\delta V_{N} (R)+i\delta W_{N} (R)$ is the long-range nuclear dynamical polarization potential (or NDPP) which accounts for the nuclear breakup and transfer reactions. We note that it is possible to split the NDPP into two parts, one for nuclear breakup and the other for transfer. To a first approximation these may employ the same diffuseness and radius, just the strengths being varied to fit the corresponding cross section data if these are available.
In reactions induced by weakly-bound projectiles with light targets we may ignore the CDPP. For more complex reactions, other components can be added.

%=========================================================

\subsection{\label{sec:cs} Cross sections}
%==========================================================
%=========================================================

\subsubsection{\label{sec:cs1} Partial and total reaction cross sections}
%==========================================================
Using the continuity equation, the total reaction cross section can be calculated from the imaginary potential as
\begin{equation}
\label{eq:j4}
\sigma_{\mathrm{Reac}} = - \frac{2}{\hbar \upsilon} \left\langle \psi | W | \psi  \right\rangle = - \frac{2}{\hbar \upsilon} \int d^3 R | \psi (R) |^2 W(R)
\end{equation}
where $\upsilon$ is the asymptotic relative velocity and $\psi$ is the usual distorted wave function that satisfies the Schr\"{o}dinger equation with the full optical model potential $U(R)=V(R)+iW(R)$.

Similarly, the direct reaction and fusion cross sections can be calculated within the extended optical model using the imaginary surface type direct-reaction and volume type fusion potentials, respectively~\cite{Hon89,Uda89,Hon89,Kim90,So05}. Here we have three contributions to the absorption: fusion, direct nuclear, and direct Coulomb, so the total fusion and direct cross sections are calculated as
\begin{eqnarray}
\label{eq:j5}
\sigma_{\mathrm{Reac}} &=&\sigma_{F}+\sigma_{DN}+\sigma_{DC}  \nonumber \\
&=&- \frac{2}{\hbar \upsilon} \left\langle \psi | W_{N} (R)+ \delta W_N (R)+\delta W_C (R)| \psi  \right\rangle 
%\nonumber \\ &=& - \frac{2}{\hbar \upsilon} \int d^3 r | \psi (R) |^2 (W_{N} (R)+ \delta W_N (R)+\delta W_C (R))
\end{eqnarray}
and then
\begin{equation}
\sigma_{i} = \frac {2}{\hbar v} \left\langle \psi |W_{\it i}(R)|\psi \right\rangle
\hspace{.5in} (i=DN, DC, \;\mbox{or}\;F).
\end{equation}
Note that DN refers to direct nuclear reactions like transfer and nuclear breakup. DC refers to the direct Coulomb breakup.
In terms of the partial-wave radial wave functions $\chi_{\ell} (R)$, the complete wave function, $\psi (\mathbf{R})=\psi (R,\theta)$, of the Schr\"{o}dinger equation can be expanded as
\begin{equation}
\label{eq:w1}
\psi (\mathbf{R}) =  \frac{1}{k R} \sum_{{\ell}=1}^{\infty} (2{\ell}+1)i^\ell \chi_\ell (R) P_{\ell} (cos(\theta))
\end{equation}
where $P_{\ell} (cos(\theta))$ are Legendre functions and satisfy the orthogonality relation
\begin{equation}
\label{eq:p1}
\int_{-1}^{1} dcos(\theta)P_{\ell} (cos(\theta))P_{\acute{{\ell}}}(cos(\theta)) =  \frac{2}{2{\ell}+1} \delta_{{\ell} \acute{{\ell}}}
\end{equation}
and then
\begin{eqnarray}
\label{eq:sigr3}
%\sigma_{i} = - \frac{2}{\hbar \upsilon} \int dR 4\pi r^2 W_{i}(R) \sum_{{\ell}=1}^{\infty} (2{\ell}+1) \left| \frac{\chi_{\ell} (R)}{k r}\right| ^2
\sigma_{i} & = & \sum_{\ell} \sigma_{i;\ell} =- \frac{2}{\hbar \upsilon} \frac{4\pi}{k^2} \sum_{{\ell}=1}^{\infty} (2{\ell}+1)\int dR |\chi_{\ell} (R)|^2 W_{i}(R) \nonumber \\
& = & \frac{\pi}{k^2}\sum_{\ell}(2\ell+1)T_{i,\ell},
\end{eqnarray}
where the transmission coefficient ($T_{i,\ell}$) is given by
\begin{equation}
T_{i,\ell}  =  \frac{8}{\hbar \upsilon} \int_{0}^{\infty} | \chi_{\ell}(R) |^{2} W_i(R) dR.
\end{equation}
Thus, for a given shape the depths of the DN and F imaginary potentials may be fixed by fitting the corresponding
cross section data if these are available, although the strength of the fusion imaginary potential is usually held fixed.
In practice, the strengths of the real and imaginary parts of the DN potential are usually fixed by fitting the elastic scattering data and these are the only adjustable parameters of the present model.

%=========================================================

\subsubsection{\label{sec:cs2} Calculation of angular distributions}
%==========================================================
In the semi-classical approximation~\cite{bass,sat1,mott,ford}, the trajectory impact parameter $b$ and orbital angular momentum $\ell$ are related to the scattering angle in the center-of-mass frame $\theta_{\textrm{c.m.}}$ by
\begin{equation}
b=\frac{\ell}{k}=\frac{D_{0}}{2}\mbox{cot}\frac{\theta_{\textrm{c.m.}}}{2},
\end{equation}
where $D_{0}=\frac{Z_{P}Z_{T}e^{2}}{E_\mathrm{c.m.}}$ is the distance of closest approach in a head-on collision, $k=\sqrt{2\mu E_{\mathrm{c.m}}}/\hbar$ is the wave number, $E_\mathrm{c.m.}$ is the incident energy in the center-of-mass system, and $Z_P$ and $Z_T$ the charges of the projectile and target ions, respectively. By treating $\ell$ as a continuous variable and assuming that $\frac{d\sigma_{i}(\ell)}{d\ell} =\sigma_{i;\ell}$ ~\cite{Kim02a},
the angular distribution of the cross section for each potential is given as ~\cite{Kim02a}
\begin{eqnarray}
\frac{d\sigma_{i}(\ell)}{d\Omega} & = &
\frac{1}{2\pi \mbox{sin}(\theta_{\textrm{c.m.}})}\frac{d\ell}{ d\theta_{\textrm{c.m.}}}
\frac{d\sigma_{i}(\ell)}{d\ell} \nonumber \\
& = &\frac{kD_{0}}{16\pi} \frac{1}{\mbox{cos}\left(\frac{\theta_{\textrm{c.m.}}}{2}\right)\mbox{sin}^{3}\left(\frac{\theta_{\textrm{c.m.}}}{2}\right)}
\sigma_{i;\ell}.
\end{eqnarray}
The angular distribution of the total transfer plus breakup cross section then comes from the direct nuclear and Coulomb contributions and is written as ~\cite{Kim02a,So16}:
\begin{equation}
\frac{d\sigma_{\mathrm{BU}}}{d\Omega}=\frac{kD_{0}}{16\pi} \frac{1}{\mbox{cos}\left(\frac{\theta_{\textrm{c.m.}}}{2}\right)\mbox{sin}^{3}\left(\frac{\theta_{\textrm{c.m.}}}{2}\right)}\sigma_{\mathrm{BU}; l}
\label{cross-section}
\end{equation}
with
\begin{equation}
\sigma_{\mathrm{BU}; l} = \frac{\pi}{k}(2l + 1)~\frac{8}{\hbar \upsilon} \int_{0}^{\infty} | \chi_{\ell}(R) |^{2} [\delta W_{C}(R) +\delta W_{N}(R) ]dR.
\label{T-br}
\end{equation}
To calculate the differential cross section for the elastic scattering, we start from the Rutherford cross section $\sigma_{\mathrm{R}}=\sum_{\ell}\sigma_{\mathrm{R}; \ell}$, where \cite{Kim02a}
\begin{equation}
\label{eq:sigRuthl}
\sigma_{\mathrm{R}; \ell} =\frac{d\sigma_{\mathrm{R}}(\ell)}{d\Omega}=\frac{\pi}{k^2} (2\ell+1),
\end{equation}
and then
\begin{equation}
	\frac{d\sigma_{\mathrm{R}}}{d\Omega}=\frac{kD_{0}}{16\pi} \frac{1}{\mbox{cos}\left(\frac{\theta_{\textrm{c.m.}}}{2}\right)\mbox{sin}^{3}\left(\frac{\theta_{\textrm{c.m.}}}{2}\right)}\frac{\pi}{k^2}(2\ell+1).
	\label{eq:sigRuth}
\end{equation}
The elastic transmission for each $\ell$ can be obtained as $T_{El,\ell} = 1 -\sum_i T_{i,\ell}=1-(T_{F,,\ell}+T_{DC,\ell}+T_{DN,\ell})$ and then
\begin{equation}
	\frac{d\sigma_{\mathrm{El}}}{d\Omega}=\frac{d\sigma_{\mathrm{R}}}{d\Omega}-\sum_i \frac{d\sigma_{\mathrm{i}}}{d\Omega}
	\label{eq:sigEl}
\end{equation}
where i = F, DC, DN. 
Here it is convenient to work with the ratio of the differential cross section to the Rutherford cross section
\begin{eqnarray}
P_i& = &\frac{d\sigma_{\mathrm{i}}}{d\Omega} \bigg/ \frac{d\sigma_{\mathrm{R}}}{d\Omega} \nonumber \\
	& = &\frac{2\ell+1}{kD_{0}} \mbox{tan}\left(\frac{\theta_{\textrm{c.m.}}}{2} \right)\approx T_{i,\ell}
\end{eqnarray}
where
\begin{equation}
	P_{El}+P_{F}+P_{DC}+P_{DN}\approx 1.
\end{equation}
%============================================

\section{\label{sec:app} Application to the $^{6}$H\lowercase{e}+$^{209}$B\lowercase{i} system}
%============================================
\begin{figure}
	\centering
	\includegraphics[width=0.45\textwidth,clip=]{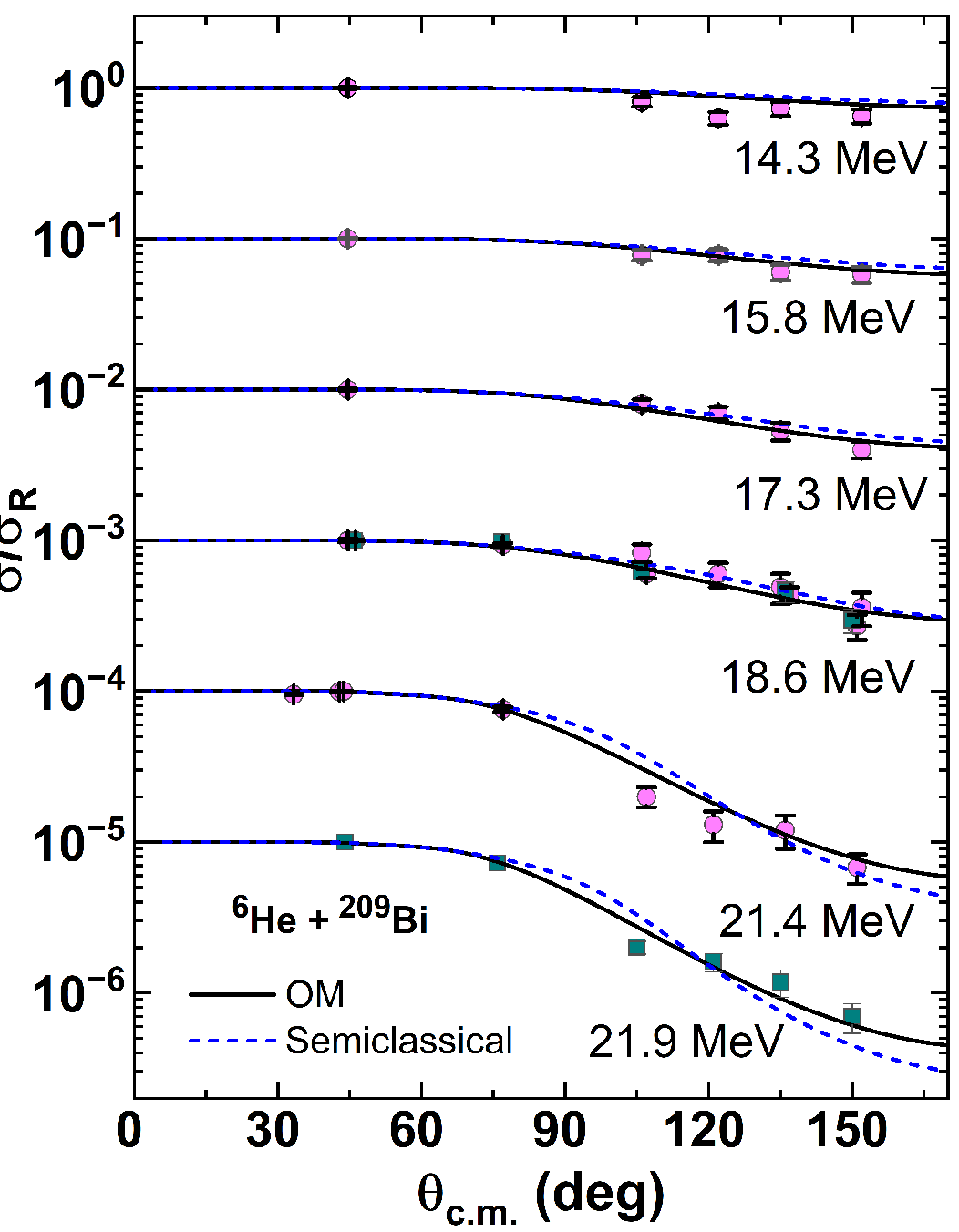}
	\caption{\label{fig:de} $^{6}$He+$^{209}$Bi elastic scattering angular distributions calculated using the optical model (solid lines) and the semi-classical model (dashed lines) compared with the data from Refs.~\cite{Agu01,Agu00}. The c.m. energies corresponding to each distribution are given.}
\end{figure}
\begin{table}
	\centering
	\caption{\label{tab:cs} OM calculations of cross sections for $^{6}$He+$^{209}$Bi. }
	\begin{tabularx}{\linewidth}{@{\extracolsep{4pt}}cllccclc}
		\hline\noalign{\smallskip}
		$E_{\bf c.m.}$ &$V_{L}$&$W_{L}$ &  $\sigma_{F}$&$\sigma^{C}_{\mathrm{bu}}$&$\sigma_{\mathrm{DN}}$ &$\sigma_{\alpha}$ &$\sigma_{\mathrm{Reac}}$  \\
		(MeV)&(MeV)&(MeV) &  (mb)&(mb)&(mb) &(mb) &(mb)  \\
		\hline\noalign{\smallskip}
	14.3	&1.7	&0.61	&0.4	&51	    &162	&213	&214 \\
	15.8	&1.0	&0.52	&4.0	&87	    &292	&379	&383 \\
	17.3	&0.61	&0.45	&22.9	&127	&435	&561	&584 \\
	18.6	&0.32	&0.36	&71.9	&161	&489	&650	&722 \\
	21.4	&0.14	&0.29	&339	&219	&594	&813	&1152 \\
	21.9	&0.08	&0.27	&398	&227	&578	&805	&1202 \\
		\hline\noalign{\smallskip}	 	\end{tabularx}
	%\end{ruledtabular}
\end{table}
%-------------------------------------------------------------
\begin{figure}
	\centering
		\includegraphics[width=0.45\textwidth,clip=]{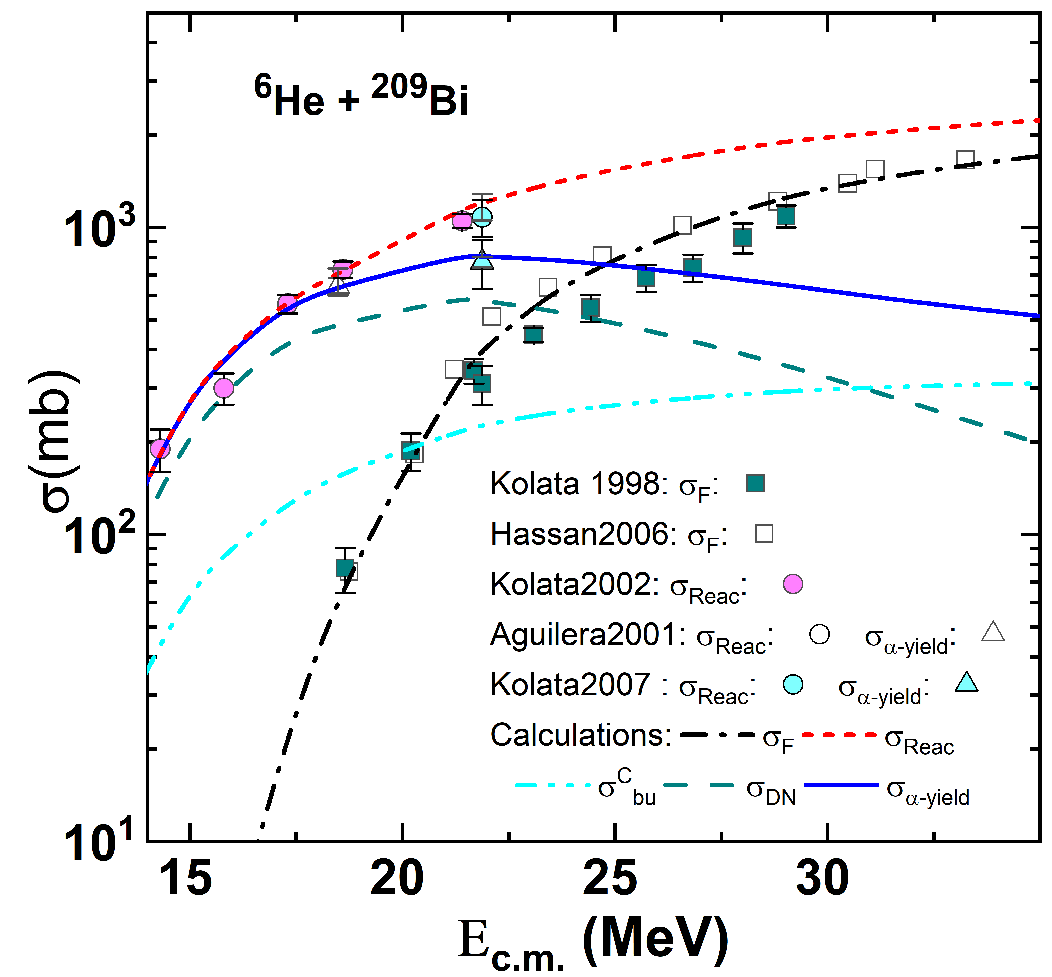}
	\caption{\label{fig:sig}  Fusion ($\sigma_{F}$), Coulomb breakup ($\sigma^{C}_{\mathrm{bu}}$), direct nuclear ($\sigma_{\mathrm{DN}}$), total direct ($\sigma_{\alpha-\mathrm{yield}}$) and total reaction ($\sigma_{\mathrm{Reac}}$) cross sections calculated with the $^{6}$He+$^{209}$Bi optical potential from the present work compared with the experimental data \cite{Kol98,Has06,Kol02,Agu00}.}
\end{figure}

\begin{figure}
	\centering
	\includegraphics[width=0.45\textwidth,clip=]{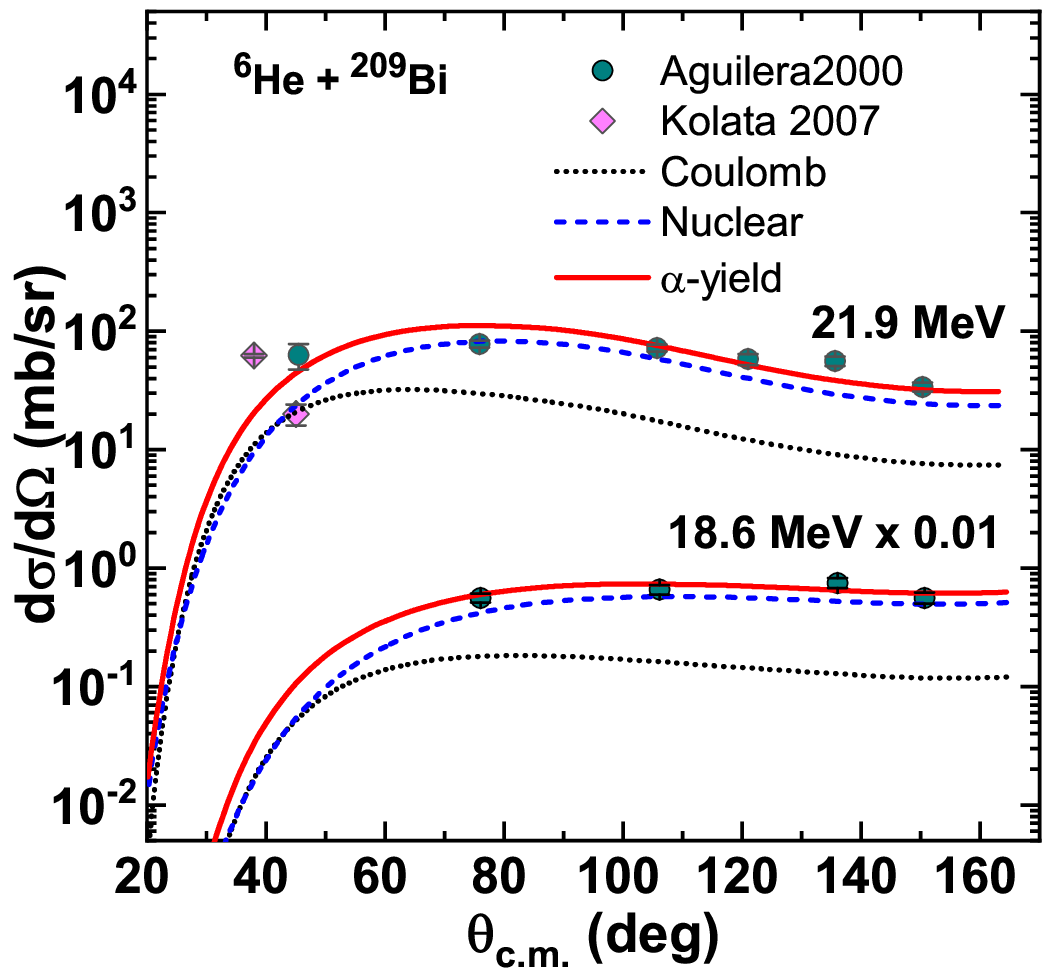}
	\caption{\label{fig:sigbu} Calculated direct reaction (transfer plus breakup) angular distributions for the
$^{6}$He $^{209}$Bi system at c.m. incident energies of 18.6 and 21.9 MeV compared with the experimental data. The
dotted and dashed curves denote the Coulomb breakup and the direct nuclear contributions, respectively. The circles denote the total $\alpha$ yield angular distributions of Aguilera \textit{et al.} \cite{Agu00}. The diamonds denote the total breakup cross section data of Kolata \textit{et al.} \cite{Kol07}.}
\end{figure}
%-------------------------------------------------------------
We apply the above methodology to the $^{6}$He+$^{209}$Bi system.
This system has been studied many times before, see for example Refs.~\cite{Kim02b,Gar07,Yan14,Fer23,Yan17,Mat06}, since
it has the most complete data set of any system involving a weakly-bound exotic projectile. It thus provides a severe
test of the ability of the present formalism to describe a wide body of data with only two adjustable parameters. 
Remarkably high yields for $\alpha$-particle emission have been observed in studies \cite{Agu00,Agu01,Kol07} of the $^{6}$He+$^{209}$Bi interaction at energies close to the Coulomb barrier. They have been shown to arise from one-neutron transfer \cite{Byc04}, two-neutron transfer \cite{You05}, and projectile breakup \cite{Kol07}. The transfer accounts for nearly 75\% of the total $\alpha$-particle yield \cite{Rus05,Kol07}.

The calculations are carried out using the optical model framework with the total optical potential of Eq.~\ref{eq:OP1}. 
In the SPP, the density of $^{209}$Bi is given by a two-parameter Fermi distribution obtained by fitting the appropriate Hartree-Fock density.
The density distribution of $^{6}$He is of Gaussian-oscillator form, $\rho =\rho_\mathrm{core} + \rho_\mathrm{valence}$, where the core density is usually taken as a single-parameter Gaussian and the density of the valence nucleon(s) is assumed to have a 1$p$-shell harmonic oscillator distribution. Its parameters were obtained by fitting the measured proton elastic scattering cross sections at high incident energies using the Glauber multiple scattering theory \cite{Xua15}.
In the CDPP, $^{6}$He is described within the $^{4}$He + $2n$ cluster model of Moro \textit{et al.\/} \cite{Mor07} with a separation energy of $\varepsilon_0=1.6$ MeV that simulates the wave functions of realistic three-body calculations and gives a very good description of the elastic scattering data for several reactions induced by $^{6}$He \cite{Mor07}.
Accordingly, the diffuseness of the NDPP is given as 1.565 fm as noted in Sec. \ref{sec:NDPP}. 

We analyzed the elastic scattering angular distributions for the $^{6}$He+$^{209}$Bi system measured at energies around the Coulomb barrier, namely, at c.m.\ energies of 14.3, 15.8, 17.4, 18.6, 21.4, and 21.9 MeV \cite{Agu01,Agu00}.
The results of the optical-model calculations are presented in Fig.~\ref{fig:de}.
To check whether the semiclassical approximation is sufficiently accurate, the angular distributions of the elastic 
cross sections calculated using Eqs.~(\ref{eq:sigRuth}) and (\ref{eq:sigEl}) are also shown in Fig.~\ref{fig:de}; 
they are very close to the optical model results.
At the same time the cross sections were calculated for the fusion, Coulomb breakup, other direct nuclear yields (nuclear breakup and transfer), and the total direct yield (the sum of the direct channels: Coulomb breakup and nuclear breakup plus transfer, corresponding to the measured inclusive $\alpha$ yield). Note that the DN component will in principle also include a contribution from inelastic excitation of the target, but for this system it is completely negligible. These partial cross sections and the total reaction cross section are listed in Table \ref{tab:cs} and presented in Fig.~\ref{fig:sig}. The angular distributions of the total direct yield are compared with the measured inclusive $\alpha$ yields at $E_\mathrm{c.m.} = 18.6$ and 21.9 MeV \cite{Agu00} in Fig.~\ref{fig:sigbu}.
 Note that the calculated angular distributions are in the center-of-mass frame whereas the measured
inclusive $\alpha$ yields should, due to their inclusive nature, in principle be in the laboratory frame.
Thus, strictly speaking a direct comparison is not possible. However, the
data of Ref.~\cite{Agu00} are plotted therein as a function of $\theta_{\rm c.m.}$; it is not clear whether this is simply
a typographical error or the transformation was made assuming a specific reaction mechanism. Nevertheless, given the
large mass asymmetry of the scattering system the difference between laboratory and center-of-mass frames should be relatively small.
The breakup data of Ref.~\cite{Kol07} are the result of a coincidence measurement and were transformed to the 
reconstructed $^6$He$^*$ center-of-mass frame. They may thus be compared directly with the Coulomb breakup angular
distribution denoted by the dotted curve on Fig.~\ref{fig:sigbu}.

The obtained $V_{L}$ and $W_{L}$ values of the NDPP (the only free parameters in our optical potential) are listed in Table \ref{tab:cs}. They have a systematic behavior as a function of energy which may be parameterized in an exponential form as $V_{L}=315.3\exp(-E_\mathrm{c.m.}/2.74)$ and $W_{L}=2.89\exp(-E_\mathrm{c.m.}/9.19)$.
Since there are fusion cross section data at energies larger than 21.9 MeV, the calculations were extended to include c.m.\ energies larger than 22 MeV using the obtained energy dependence of $V_{L}$ and $W_{L}$. Thus at these energies we do not have any free parameter in our potential.
All the data are simultaneously well reproduced. However, it is possible to obtain almost as good fits with $V_{L}$ fixed at zero, i.e.\ with a purely imaginary NDPP and thus only one adjustable parameter, although allowing $V_{L}$ to vary does give improved agreement, in particular with the total reaction and $\alpha-\mathrm{yield}$ cross sections.
We note here that in our discussion we refer to the direct nuclear part of the optical potential as consisting of the combined effects of transfer and nuclear breakup.
However, in this system the Coulomb breakup is dominant and much more important than the nuclear, so we do not consider the nuclear breakup separately since the transfer cross section is the dominant contributor to the direct nuclear (DN) cross section.

\begin{figure}
	\centering
	\includegraphics[width=0.45\textwidth,clip=]{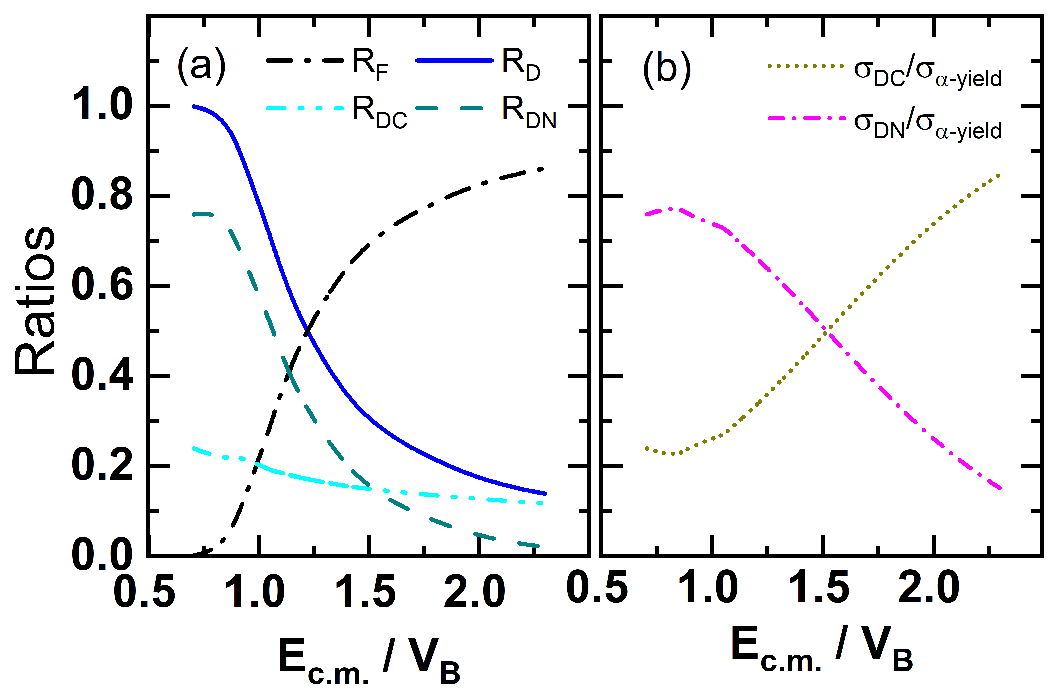}
	\caption{\label{fig:Ratios} (a) Ratios of fusion ($R_F$), Coulomb direct (Coulomb breakup $R_{DC}$), nuclear direct (nuclear breakup and transfer $R_{DN}$), and total direct or $\alpha$-yield ($R_{D}$) cross sections to  total reaction cross sections for the $^{6}$He+$^{209}$Bi system. (b) Ratios of the Coulomb and direct nuclear contributions to the total direct cross section ($\sigma_{\alpha-\mathrm{yield}}$).}
\end{figure}

Further to elucidate the behavior of the different contributions to the total cross section, we define the ratios $R_i=\sigma_i/\sigma_{\mathrm{Reac}}$ where $i= F$, $DC$, $DN$, $D$ and $F$, $DC$, $DN$, $D$ refer to the fusion, Coulomb direct (Coulomb breakup), nuclear direct (nuclear breakup and transfer), and total direct ($\alpha$-yield), respectively. These ratios are shown in Fig.~\ref{fig:Ratios} (a) as a function of $E_\mathrm{c.m}/V_B$ where $V_B$ is the Coulomb barrier (about 20.3 MeV \cite{Agu01,Kol98}). We see that below the barrier the total direct contribution is dominant, whereas above the barrier fusion is the dominant contributor to the total cross section. The nuclear contribution to the $\alpha$-yield cross section (transfer and nuclear breakup) is much larger than the Coulomb contribution below the Coulomb barrier. This rule is reversed for energies higher than 32 MeV ($E_\mathrm{c.m}/V_B \approx 1.5$) and the Coulomb breakup eventually becomes dominant at high enough energies. This is illustrated more clearly in Fig.~~~\ref{fig:Ratios} (b) where the ratios of the Coulomb direct and nuclear direct cross sections to the $\alpha-\mathrm{yield}$ cross section are plotted.

At $E_\mathrm{c.m} \approx 22$ MeV the calculated Coulomb breakup is 227 mb which is close to the experimental value of 205(65) mb \cite{Kol07} or the calculation of Ref.~\cite{Rus05}, 218 mb.
The calculated DN cross section is about 578 mb which is in agreement with the measured transfer cross section of 565 mb \cite{Kol07}. The $\alpha$ yield is about 805 mb, which is in agreement with the measured values 770(140) mb \cite{Kol07} and 773(31) mb \cite{Agu00,Agu01}.
The calculated total reaction cross section, the sum of the fusion and direct reaction yields, is in good agreement with the experimental values of 1080(148) mb \cite{Kol07} or 1170(150) mb \cite{Agu00,Agu01} and the calculation of Rusek \cite{Rus05}, 1182 mb.

At 18.6 MeV, our calculated value for the $\alpha$-yield cross section of 650 mb is close to the measured one of 643(42) mb \cite{Agu00}.
The direct nuclear contribution is about 75\% of the $\alpha$-yield cross section at all the energies considered here, which is the same ratio deduced from previous measurements and calculations \cite{Byc04,You05,Rus05,Kol07} assuming that transfer is the main component of the direct nuclear processes.

\begin{figure*}
	\centering
		\includegraphics[width=0.32\textwidth,clip=]{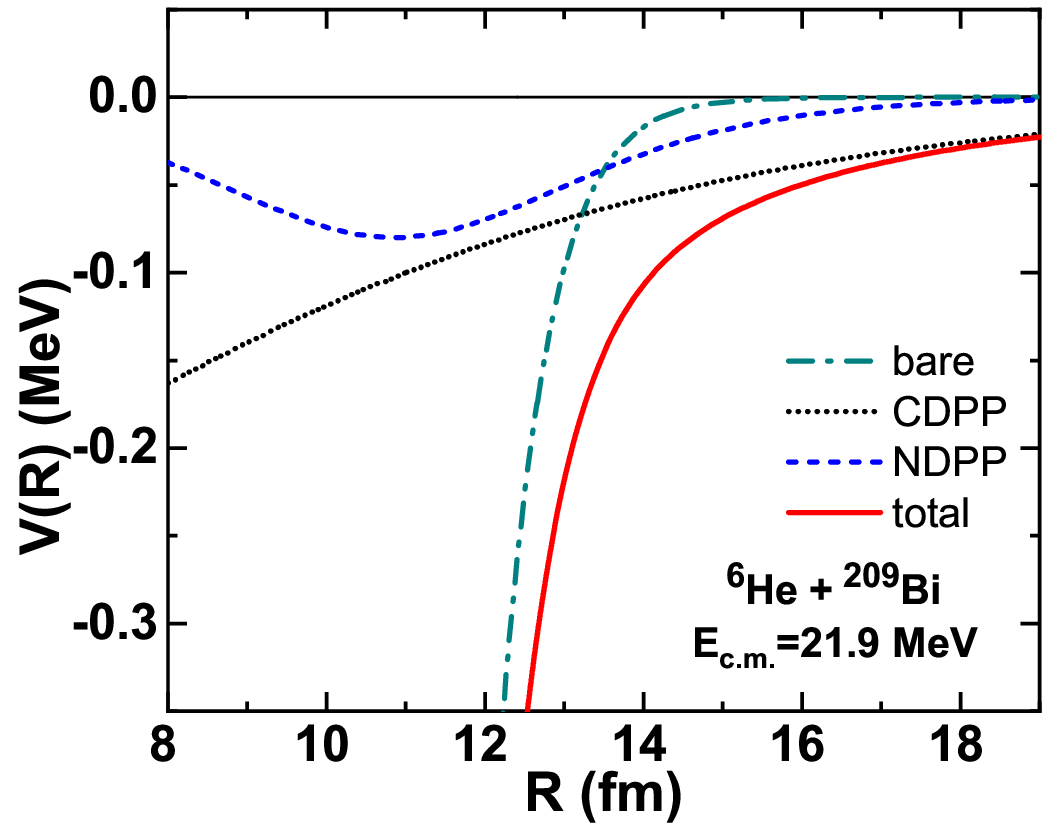}
		\includegraphics[width=0.32\textwidth,clip=]{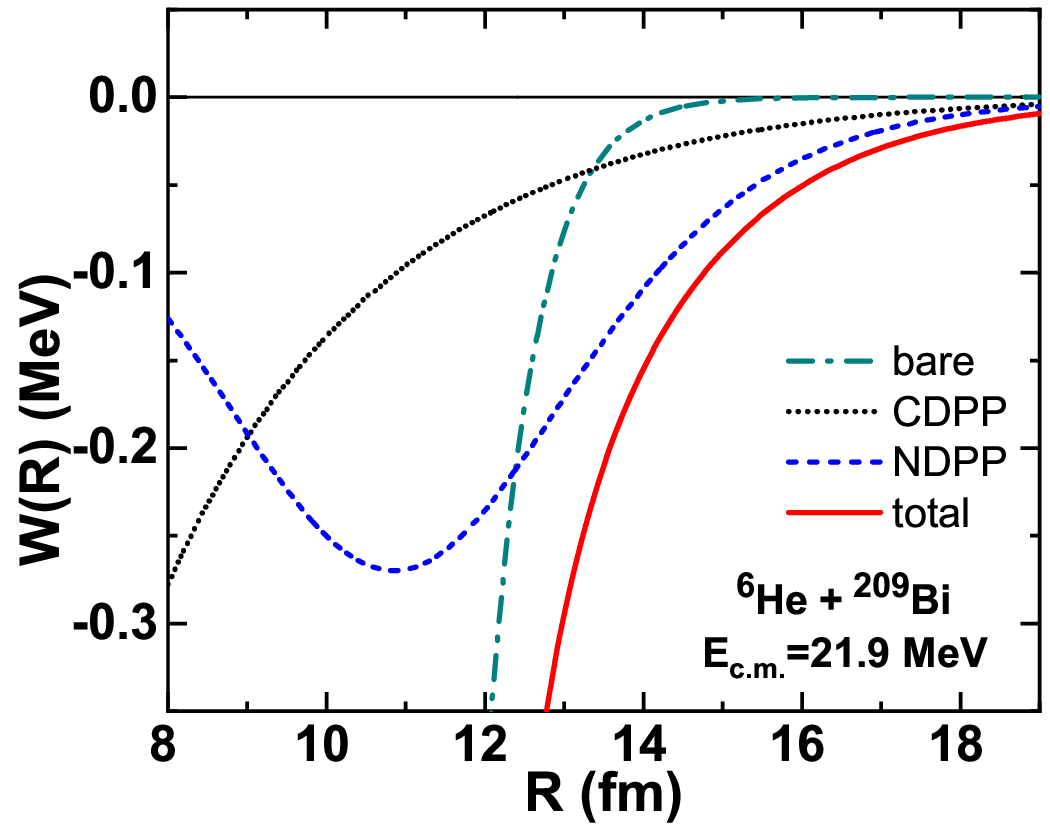}
		\includegraphics[width=0.32\textwidth,clip=]{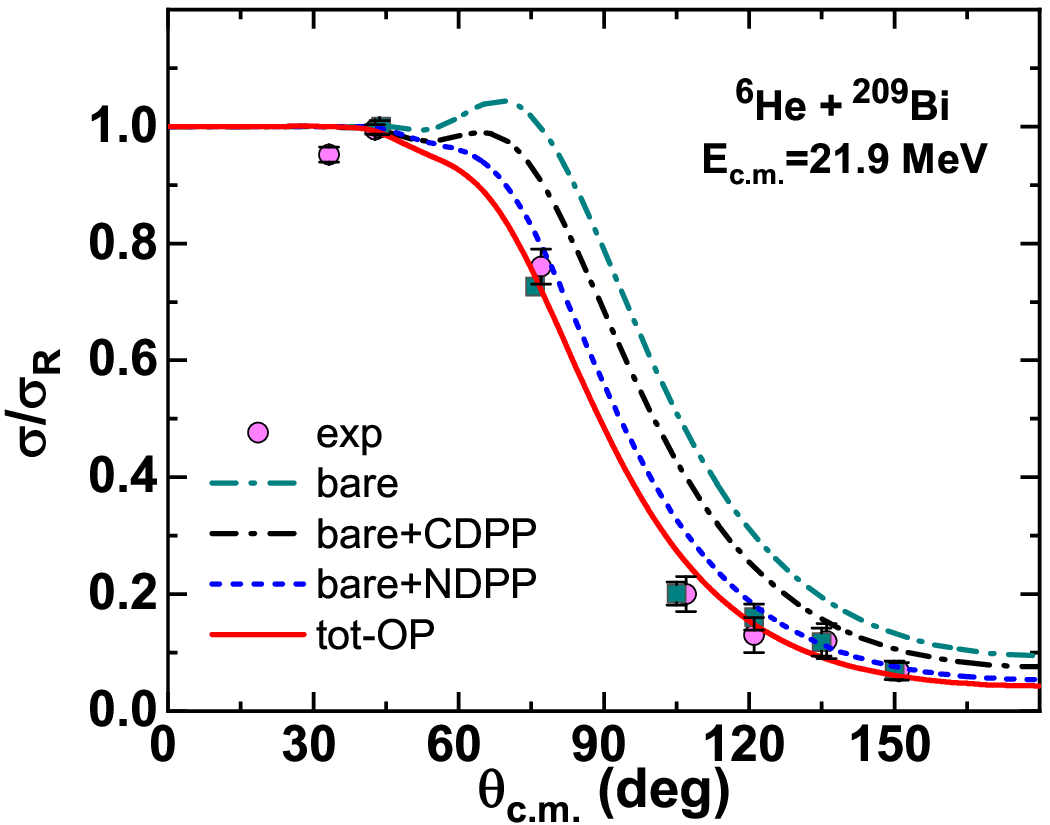}
	\caption{\label{fig:pot} (a) and (b) The real and imaginary parts of the bare, CDPP and NDPP potentials at 21.9 MeV. Note that the bare and CDPP potentials do not depend on the incident energy.
		(c) Calculated angular distributions for $^{6}$He+$^{209}$Bi elastic scattering compared to the data at $E_\mathrm{c.m.} = 21.9$ MeV \cite{Agu01,Agu00} showing the influence of the CDPP and NDPP.
		 }
\end{figure*}

Figure \ref{fig:pot} shows the bare, CDPP, NDPP, and the total potentials used to calculate the cross sections for the $^{6}$He+$^{209}$Bi system at $E_\mathrm{c.m.} = 21.9$ MeV. The real CDPP has the longest range due to the polarization of the $^{6}$He projectile and the imaginary CDPP and NDPP, which account for the loss of flux due to the other direct reaction processes, are also of much longer range than the bare potential. Figure \ref{fig:pot} (c) shows that adding the CDPP alone to the bare potential cannot reproduce the data and the long-range NDPP is needed to account for the full deviation from the Rutherford cross section. 

%===============
	\begin{figure}
		\centering
		\includegraphics[width=0.4\textwidth,clip=]{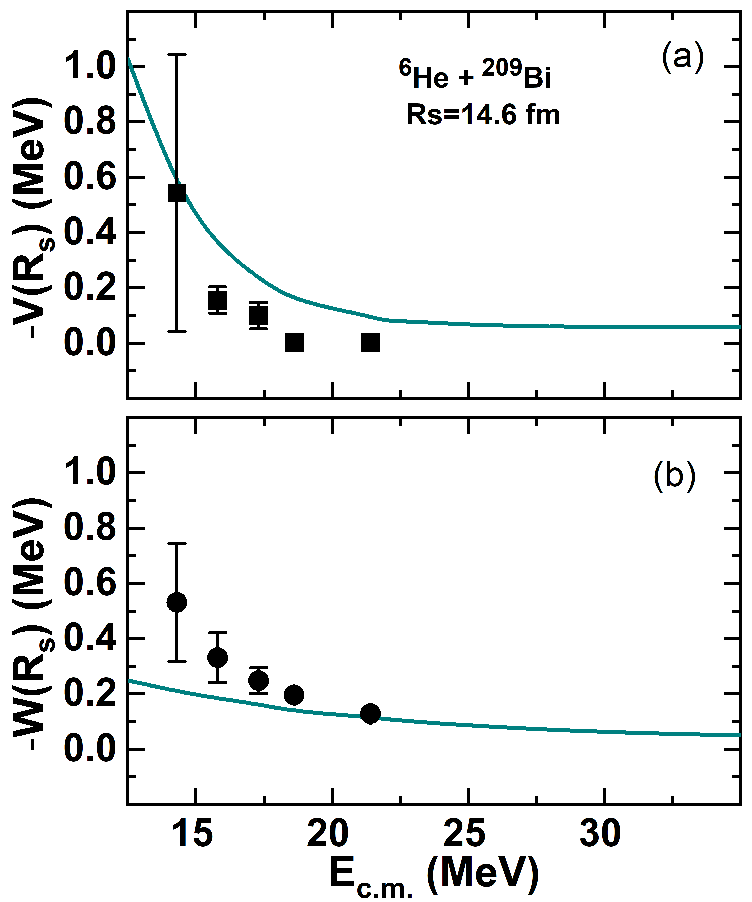}
		\caption{\label{fig:Rs} Energy dependence of the real (a) and imaginary (b) parts of the total optical potential (\ref{eq:OP1}) at the sensitivity radius of 14.6 fm for $^{6}$He+$^{209}$Bi system in comparison with the results of previous optical model fits \cite{Gar07}.}
	\end{figure}
	Figure \ref{fig:Rs} shows the energy dependence of the real (a) and imaginary (b) parts of the total optical potential (\ref{eq:OP1}) at the sensitivity radius of 14.6 fm \cite{Gar07} compared with the corresponding values from the optical model fits of Ref.~\cite{Gar07}. It is clear that both the real and imaginary potentials have long ranges even below the Coulomb barrier. The largest contributor to the real part at the sensitivity radius is the CDPP, whereas the imaginary NDPP is the largest contributor to the imaginary part below the Coulomb barrier. This is consistent with the observed very large cross sections for the transfer of one or two neutrons at these low energies. 
At higher energies, greater than 32 MeV, the imaginary CDPP becomes larger than the imaginary NDPP and Coulomb breakup makes the largest contribution to the direct cross section, see Fig.~\ref{fig:Ratios} (b).
%============================================

\section{\label{sec:summary} Summary and conclusions}
%============================================
In summary, a simultaneous analysis has been carried out of cross section data for elastic scattering, fusion, Coulomb breakup and other direct nuclear channels of the $^{6}$He+$^{209}$Bi system within the framework of the extended optical model \cite{Uda84,Uda85,Uda89,Hon89,Kim90}. The optical potential used consisted of a short-range bare nuclear potential (volume type), long-range NDPP (surface type), and CDPP. The bare potential was calculated using the SPP prescription \cite{Cha02} and the CDPP according to a recent formalism \cite{Mar21,Mar22}. The NDPP was of Woods-Saxon derivative form; however, guided by semiclassical theory and the observation that the results are essentially sensitive to just the tail of this potential, the radius and diffuseness parameters could be fixed leaving the real and imaginary depths of the NDPP, $V_L$ and $W_L$, as the only free parameters of the potential adjusted to fit the elastic scattering angular distribution and reaction cross section data.  
The angular distribution of the total direct cross section derived from the summation of the DN (transfer and nuclear breakup) and DC (Coulomb breakup) channels was also compared with the measured inclusive $\alpha$ production angular distributions at $E_\mathrm{c.m.} = 18.6$ and 21.9 MeV \cite{Agu00}. All the calculated cross sections are in good agreement with the data.

It was found that $V_L$ and $W_L$ exhibited a simple exponential dependence on the incident energy, enabling calculation of the fusion cross section for energies where no elastic scattering data exist but where the fusion has been measured. These predictions---the values of $V_L$ and $W_L$ were fixed following the systematics---were in good agreement with the data. Thus the methodology also has some predictive power via extrapolation into regions where there are no existing data. The success of the model is at least in part due to the applicability to the system under study of the semiclassical concepts employed. Also, the available observables are such that they appear to be relatively insensitive to interference terms between the Coulomb and nuclear breakup mechanisms which cannot be handled within the present formalism. With these limitations, the present model has the advantage of being able to describe well a large body of data over a range of near-barrier energies with only two free parameters. As such it should prove of use in planning experiments and also as a source of ``pseudo data'' that may be used to help validate more sophisticated models.
%\bigskip
\section*{Acknowledgments}
Thanks to Dr.\ D. K. Sharp for input into the preparation of the manuscript. This work was funded by the Council for At-Risk Academics (Cara) within the Cara Fellowship Programme. This work was partially supported by the British Academy within the British Academy/Cara/Leverhulme Researchers at Risk Research Support Grants Programme under grant number LTRSF/100141.
%%%%%%%%%%%%%%%%%%%%%%%%%%%%%%%%%%%%%%%%%%%%%%%%%%%%%%%%%%%%%%%%%%%%%%%%%
%%%%%%%%%%%%%%%%%%%%%%%%%%%%%%%%%%%%%%%%%%%%%%%%%%%%%%%%%%%%%%%%%%%%%%%%%


\begin{thebibliography}{}
	%\href{http://www}{Something }  \url{http://www...com}
	%%%%%%% introduction
	\bibitem{Kee09} N. Keeley, N. Alamanos, K. W. Kemper, and K. Rusek,
	\href{https://doi.org/10.1016/j.ppnp.2009.05.003} {Prog. Part. Nucl. Phys. 63, 396 (2009).}
	\bibitem{Can15} L. F. Canto, P. R. S. Gomes, R. Donangelo, J. Lubian, and M. S. Hussein,
	\href{https://doi.org/10.1016/j.physrep.2015.08.001} {Phys. Rep. 596, 1 (2015).}
	\bibitem{Kol16} J. J. Kolata, V. Guimar\~{a}es, and E. F. Aguilera,
	\href{https://doi.org/10.1140/epja/i2016-16123-1} {Eur. Phys. J. A 52, 123 (2016).}
	\bibitem{Uda84} T. Udagawa and T. Tamura, \href{https://doi.org/10.1103/PhysRevC.29.1922}{Phys. Rev. C  {\bf 29}, 1922 (1984).}
	\bibitem{Uda85} T. Udagawa, B. T. Kim, and T. Tamura \href{https://doi.org/10.1103/PhysRevC.32.124}{Phys. Rev. C \textbf{32}, 124 (1985).}
	%T. Udagawa and T. Tamura, ibid. 29, 1922 (1984);
	\bibitem{Hon89} S.-W. Hong, T. Udagawa, and T. Tamura, 
	\href{https://doi.org/10.1016/0375-9474(89)90582-4}{Nucl. Phys. A \textbf{491}, 492 (1989).}
	\bibitem{Uda89} T. Udagawa, T. Tamura, and B. T. Kim, \href{https://doi.org/10.1103/PhysRevC.39.1840}{Phys.  Rev. C  {\bf 39}, 1840 (1989).}
	\bibitem{Kim90} B. T. Kim, M. Naito, and T. Udagawa,
	\href{https://doi.org/10.1016/0370-2693(90)90453-D}{Phys. Lett. B {\bf 237}, 19 (1990).}
	\bibitem{Kim02a} B. T. Kim, W. Y. So, S. W. Hong and T. Udagawa,
	\href{https://doi.org/10.1103/PhysRevC.65.044607}{Phys. Rev. C {\bf 65}, 044607 (2002).}
	\bibitem{Kim02b} B. T. Kim, W. Y. So, S. W. Hong and T. Udagawa,
	\href{https://doi.org/10.1103/PhysRevC.65.044616}{Phys. Rev. C {\bf 65}, 044616 (2002).}
	\bibitem{So07} W. Y. So, T. Udagawa, K. S. Kim, S. W. Hong, and B. T. Kim,
	\href{https://doi.org/10.1103/PhysRevC.75.024610}{Phys. Rev. C \textbf{75}, 024610 (2007).}
	\bibitem{So10} W. Y. So, T. Udagawa, K. S. Kim, S. W. Hong, and B. T. Kim,
	\href{https://doi.org/10.1103/PhysRevC.81.047604}{Phys. Rev. C \textbf{81}, 047604 (2010).}
	\bibitem{Mar21} H. M. Maridi, K. Rusek, and N. Keeley, \href{https://doi.org/10.1103/PhysRevC.104.024614} {Phys. Rev.C \textbf{104}, 024614 (2021).}
	\bibitem{Mar22} H. M. Maridi, K. Rusek, and N. Keeley, \href{https://doi.org/10.1103/PhysRevC.106.054613} {Phys. Rev.C \textbf{106}, 054613 (2022).}
	\bibitem{So14} W. Y. So, K. S. Kim, K. S. Choi, and M.-K. Cheoun,
	\href{https://doi.org/10.1103/PhysRevC.90.054615}{Phys. Rev. C 90, 054615 (2014).}
	\bibitem{So15a} W. Y. So, K. S. Kim, K. S. Choi, and M.-K. Cheoun,
	\href{https://doi.org/10.1103/PhysRevC.92.014627}{Phys. Rev. C 92, 014627 (2015).}
	\bibitem{San08} A. M. S\'{a}nchez-Ben\'{\i}tez \emph{et al.},
	\href{https://doi.org/10.1016/j.nuclphysa.2008.01.030}{Nucl. Phys. A {\bf 803}, 30 (2008).}
	\bibitem{Cha02} L. C. Chamon, B. V. Carlson, L. R. Gasques, D. Pereira, C. De Conti, M. A. G. Alvarez, M. S. Hussein, M. A. C\textit{\underline{}}\^{a}ndido Ribeiro, E. S. Rossi, Jr., and C. P. Silva, 
	\href{https://doi.org/10.1103/PhysRevC.66.014610}{Phys. Rev. C \textbf{66}, 014610 (2002).}
	\bibitem{Alv03} M. A. G. Alvarez, L. C. Chamon, M. S. Hussein, D. Pereira, L. R. Gasques, E. S. Rossi, and C. P. Silva, \href{https://doi.org/10.1016/S0375-9474(03)01158-8}{Nucl. Phys. A \textbf{723}, 93 (2003).}
	\bibitem{Alv21} M. A. G. Alvarez, M. Rodr\'{\i}guez-Gallardo, J. P. Fern\'{a}ndez-Garc\'{\i}a, J. Casal, and J. A. Lay, \href{https://doi.org/10.1103/PhysRevC.103.054614}{Phys. Rev. C \textbf{103}, 054614 (2021).}
	\bibitem{Sat79} G. R. Satchler and W. G. Love, Phys. Rep. \textbf{55}, 183 (1979).
	%\bibitem{Ber77} G. Bertsch, J. Borysowicz, H. McManus, and W. G. Love, Nucl. Phys. A \textbf{284}, 399 (1977).
	\bibitem{Cre07} E. Crema, P. R. S. Gomes, and L. C. Chamon, 
	\href{https://doi.org/10.1103/PhysRevC.75.037601}{Phys. Rev. C \textbf{75}, 037601 (2007).}
	\bibitem{Ber92} C.A. Bertulani, A. Sustich, 
	\href{https://doi.org/10.1103/PhysRevC.46.2340}{Phys. Rev. C {\bf 46}, 2340 (1992).}
	\bibitem{Abr72} Handbook of Mathematical Functions, 10th ed., edited by
	M. Abramowitz and I. A. Stegun (Dover, New York, 1972).
	\bibitem{Bor07} L. Borowska, K. Terenetsky, V. Verbitsky, and S. Fritzsche,
	\href{https://doi.org/10.1103/PhysRevC.76.034606}{Phys. Rev. C \textbf{76}, 034606 (2007).}
	\bibitem{And95} M. V. Andr\'{e}s, J. G\'{o}mez-Camacho, and N. Nagarajan, 
	\href{https://doi.org/10.1016/0375-9474(94)00765-F}{Nucl. Phys. A {\bf 583}, 817 (1995).}
	\bibitem{And94} M. V. Andr\'{e}s, J. G\'{o}mez-Camacho, and N. Nagarajan, 
	\href{https://doi.org/10.1016/0375-9474(94)90806-0}{Nucl. Phys. A {\bf 579}, 237 (1994).}
	%	\bibitem{And98} M. V. Andr\'{e}s, J. G\'{o}mez-Camacho, and N. Nagarajan, 	\href{https://doi.org/10.1103/PhysRevLett.82.1387}{Phys. Rev. Lett. {\bf 82}, 1387 (1998).}	
	\bibitem{Hus94} M. S. Hussein and G. R. Satchler, \href{https://doi.org/10.1016/0375-9474(94)90732-3}{Nucl. Phys. A {\bf 567}, 165 (1994).}
	\bibitem{Nag85} M. A. Nagarajan, C. C. Mahaux, and G. R. Satchler, 
	\href{https://doi.org/10.1103/PhysRevLett.54.1136}{Phys. Rev. Lett. {\bf 54}, 1136 (1985).}
	\bibitem{Mah86} C. Mahaux, H. Ng\^{o}, and G. R. Satchler, 
	\href{https://doi.org/10.1016/0375-9474(86)90009-6}{Nucl. Phys. A {\bf 449}, 354 (1986).}
	\bibitem{Kee94} N. Keeley, S. J. Bennett, N. M. Clarke, B. R. Fulton,
	G. Tungate, P. V. Drumm, M. A. Nagarajan, and J. S.	Lilley, \href{https://doi.org/10.1016/0375-9474(94)90064-7}{Nucl. Phys. A {\bf 571}, 326 (1994).}
	\bibitem{Gar07} A. R. Garcia, J. Lubian, I. Padron, P. R. S. Gomes, T. Lacerda, V. N. Garcia, A. G\'{o}mez Camacho, and E. F. Aguilera, \href{https://doi.org/10.1103/PhysRevC.76.067603}{Phys. Rev. C \textbf{76}, 067603 (2007).}
	\bibitem{Gar08} A. R. Garcia, J. Lubian, P. R. S. Gomes, I. Padron, L. C. Chamon, D. Pereira,
	\href{https://doi.org/10.1016/j.nuclphysa.2008.03.004}{Nucl. Phys. A {\bf 806}, 146 (2008).}
	\bibitem{Fer10} J. P. Fern\'{a}ndez-Garc\'{\i}a, M. Rodr\'{\i}guez-Gallardo, M. A. G. Alvarez, and A. M. Moro, \href{https://doi.org/10.1016/j.nuclphysa.2010.03.013}{Nucl. Phys. A {\bf 840}, 19 (2010).}
	\bibitem{Kee19} N. Keeley, K. W. Kemper, I. Martel, K. Rusek, and A. M. S\'{a}nchez-Ben\'{\i}tez, 
	\href{https://doi.org/10.1103/PhysRevC.99.024603}{Phys. Rev.C \textbf{99}, 024603 (2019).}	
	\bibitem{Mac09} R. S. Mackintosh and N. Keeley, 
	\href{https://doi.org/10.1103/PhysRevC.79.014611}{Phys. Rev. C {\bf 79}, 014611 (2009).}
	\bibitem{Bon02} A. Bonaccorso and F. Carstoiub,\href{https://doi.org/10.1016/S0375-9474(02)00755-8}{ Nucl. Phys. A {\bf 706}, 322 (2002).}
	\bibitem{Has09} M. Y. M. Hassan, M. Y. H. Farag, E. H. Esmael, and H. M. Maridi, \href{https://doi.org/10.1103/PhysRevC.79.064608}{Phys. Rev. C \textbf{79}, 064608 (2009).}
	\bibitem{Mor07} A. M. Moro, K. Rusek, J. M. Arias, J. G\'{o}mez-Camacho, M. Rodriguez-Gallardo, 
	\href{https://doi.org/10.1103/PhysRevC.75.064607}{Phys. Rev. C {\bf 75}, 064607 (2007).}
	\bibitem{Pie10} A. Di Pietro, G. Randisi, V. Scuderi, L. Acosta, F. Amorini, M. J. G. Borge {\em et al.\/}
	\href{https://doi.org/10.1103/PhysRevLett.105.022701}{Phys. Rev. Lett. {\bf 105}, 022701 (2010).}
	\bibitem{Pie12} A. Di Pietro, V. Scuderi, A. M. Moro, L. Acosta, F. Amorini, M. J. G. Borge {\em et al.\/},
	\href{https://doi.org/10.1103/PhysRevC.85.054607}{Phys. Rev. C {\bf 85}, 054607 (2012).}
	\bibitem{Fes58} H. Feshbach,
	\href{https://doi.org/10.1016/0003-4916(58)90007-1}{Ann. Phys. (N.Y.) \textbf{5}, 357 (1958).}
	\bibitem{So05} W. Y. So, S. W. Hong, B. T. Kim, and T. Udagawa,
	\href{https://doi.org/10.1103/PhysRevC.69.044616}{Phys. Rev. C \textbf{69}, 064606 (2004)};
	\href{https://doi.org/10.1103/PhysRevC.72.064602}{\textbf{72}, 064602 (2005).}
	\bibitem{bass} R. Bass, {\it Nuclear Reactions with Heavy Ions}
	(Springer-Verlag, New York, 1980)
	\bibitem{sat1} G. R. Satchler, {\it Introduction to Nuclear Reactions}
	(John Wiley \& Sons, New York, 1980) p.41.
	\bibitem{mott} N. F. Mott and H. S. W. Massey, {\it The Theory of Atomic
		Collisions}, (Oxford University Press, 1965) p.97-102.
	\bibitem{ford} K. W. Ford and J. A. Wheeler, Ann. Phys. {\bf 7}, 287 (1959).
	%\bibitem{uda2} T. Udagawa and T. Tamura, Phys. Rev. C  {\bf 29}, 1922 (1984); T. Udagawa, B. T. Kim, and T. Tamura, Phys. Rev. C {\bf 32}, 124 (1985).
	\bibitem{So16} W. Y. So, K. S. Choi, M.-K. Cheoun, and K. S. Kim,
	\href{https://doi.org/10.1103/PhysRevC.93.054624}{Phys. Rev. C 93, 054624 (2016).}
	\bibitem{Fer23} J. L. Ferreira, J. Rangel , J. Lubian , and L. F. Canto, \href{https://doi.org/10.1103/PhysRevC.107.034603}{Phys. Rev. C \textbf{107}, 034603 (2023).}
	\bibitem{Yan14} L. Yang, C. J. Lin, H. M. Jia, F. Yang, Z. D. Wu, X. X. Xu, H. Q. Zhang, Z. H. Liu, P. F. Bao, L. J. Sun, and N. R. Ma, \href{https://doi.org/10.1103/PhysRevC.89.044615}{Phys. Rev. C \textbf{89}, 044615 (2014).}
	\bibitem{Yan17} L. Yang, C. J. Lin, H. M. Jia, D. X. Wang, N. R. Ma, L. J. Sun, F. Yang, X. X. Xu, Z. D. Wu, H. Q. Zhang, and Z. H. Liu, \href{https://doi.org/10.1103/PhysRevC.96.044615}{Phys. Rev. C \textbf{96}, 044615 (2017).}
	\bibitem{Mat06} T. Matsumoto, T. Egami, K. Ogata, Y. Iseri, M. Kamimura, and M. Yahiro,
	\href{https://doi.org/10.1103/PhysRevC.73.051602}{Phys. Rev. C \textbf{73}, 051602(R) (2006).}
	\bibitem{Kol07} J. J. Kolata, H. Amro, F. D. Becchetti, J. A. Brown, P. A. DeYoung, M. Hencheck {\em et al.\/}, \href{https://doi.org/10.1103/PhysRevC.75.031302}{Phys. Rev. C 75, 031302(R) (2007).}  %6He-209Bi
	\bibitem{Agu00} E. F. Aguilera, J. J. Kolata, F. M. Nunes, F. D. Becchetti, P. A. DeYoung, M. Goupell {\em et al.\/}, \href{https://doi.org/10.1103/PhysRevLett.84.5058}{Phys. Rev. Lett. 84, 5058 (2000).}  %6He-209Bi
	\bibitem{Agu01} E. F. Aguilera, J. J. Kolata, F. D. Becchetti, P. A. DeYoung, J. D. Hinnefeld, \'{A}. Horv\'{a}th, L. O. Lamm, Hye-Young Lee, D. Lizcano, E. Martinez-Quiroz, P. Mohr, T. W. O’Donnell, D. A. Roberts, and G. Rogachev, \href{https://doi.org/10.1103/PhysRevC.63.061603}{Phys. Rev. C 63, 061603(R) (2001).}  %6He-209Bi
	\bibitem{Byc04} J. P. Bychowski {\em et al.\/},
	\href{https://doi.org/10.1016/j.physletb.2004.05.058}{Phys. Lett. B596, 26 (2004).}
	\bibitem{You05} P. A. DeYoung, Patrick J. Mears, J. J. Kolata, E. F. Aguilera, F. D. Becchetti, Y. Chen {\em et al.\/},
	\href{https://doi.org/10.1103/PhysRevC.71.051601}{Phys. Rev. C 71, 051601(R) (2005).}
	\bibitem{Rus05} K. Rusek, I. Martel, J. G\'{o}mez-Camacho, A. M. Moro, and R. Raabe,
	\href{https://doi.org/10.1103/PhysRevC.72.037603}{Phys. Rev. C \textbf{72}, 037603 (2005).}
	\bibitem{Xua15} Le Xuan Chung, Oleg A. Kiselev, Dao T. Khoa, and Peter Egelhof, \href{https://doi.org/10.1103/PhysRevC.92.034608}{Phys. Rev. C \textbf{92}, 034608 (2015).}
	\bibitem{Kol98} J. J. Kolata, V. Guimar\~{a}es, D. Peterson, P. Santi, R. White-Stevens, P. A. DeYoung, G. F. Peaslee, B. Hughey, B. Atalla, M. Kern, P. L. Jolivette, J. A. Zimmerman, M. Y. Lee, F. D. Becchetti, E. F. Aguilera, E. Martinez-Quiroz, and J. D. Hinnefeld,
	\href{https://doi.org/10.1103/PhysRevLett.81.4580}{Phys. Rev. Lett. {\bf 81}, 4580 (1998).}  %6He-209Bi
	\bibitem{Has06} A. A. Hassan {\em et al.\/},
	%A. A. Hassan, S. M. Lukyanov,R. Kalpakchieva, Yu. E. Penionzhkevich, R. A. Astabatyan,I. Vinsour, Z. Dlouhy, A. A. Kulko, J. Mrazek, S. P. Lobastov, E. R. Markaryan, V. A. Maslov, N. K. Skobelev, Yu. G. Sobolev,
	%Investigation of nuclear fusion in reactions of 4,6He and 7Li with 208Pb and 209Bi nuclei
	Izv. Rossiiskoi Akademii Nauk, Ser. Fiz. {\bf 70}, 1558 (2006).
	Bull.Rus. Acad.Sci.Phys. {\bf 70}, 1785 (2006).
	\bibitem{Kol02} J. J. Kolata, \href{https://doi.org/10.1007/s10050-002-8729-x}{Eur. Phys. J. A 13, 117 (2002).}  %6He-209Bi
	%\bibitem{Jag74} C. W. de Jager, H. de Vries, and C. de Vries, At. Data Nucl. Data Tables {\bf 14}, 479 (1974).
	%\bibitem{Kee03} N. Keeley, J. M. Cook, K. W. Kemper, B. T. Roeder, W. D. Weintraub, F. Marechal, and K. Rusek,  \href{https://doi.org/10.1103/PhysRevC.68.054601}{Phys. Rev.C \textbf{68}, 054601 (2003).}
	%%%%%%%%%%%%%%%%%%%
	%	\bibitem{Yan17b} L. Yang, C. J. Lin, H. M. Jia, D. X. Wang, N. R. Ma, L. J. Sun {\em et al.\/}, \href{https://doi.org/10.1103/PhysRevLett.119.042503}{Phys. Rev. Lett. {\bf 119}, 042503 (2017).}	
	%	\bibitem{Tie91} M. A. Tiede, D. E. Trcka, and K. W. Kemper,	\href{https://doi.org/10.1103/PhysRevC.44.1698}{Phys. Rev.C \textbf{44}, 1698 (1991).}
	%	\bibitem{Pak03} A. Pakou, N. Alamanos, A. Lagoyannis, A. Gillibert, E.C. Pollacco, P. A. Assimakopoulos {\em et al.\/}, \href{https://doi.org/10.1016/S0370-2693(03)00079-0}{Phys. Lett. B {\bf 556}, 21 (2003).}
	%%%%%%%%%%
\end{thebibliography}
\end{document}